\documentclass[journal=jctcce,manuscript=article]{achemso}

\setkeys{acs}{maxauthors=0}
\SectionNumbersOn

\usepackage{amsmath}
\usepackage{amssymb}
\usepackage{amsfonts}
\usepackage{bm}
\usepackage{braket}
\usepackage{booktabs}
\usepackage{chemformula}
\usepackage{dcolumn}
\usepackage{graphicx}
\usepackage{pdfpages}
\usepackage[hidelinks]{hyperref}
\usepackage{multirow}
\usepackage{physics}
\usepackage{tabularx}
\usepackage{threeparttable}
\usepackage{xcolor}
\usepackage{xurl}

\title{Scalable Multi-GPU Multiset Tensor Network State for Quantum Dynamics}

\author{Jinjun Zeng}
\affiliation{Guangdong Basic Research Center of Excellence for Aggregate Science, School of Science and Engineering, The Chinese University of Hong Kong (Shenzhen), Shenzhen, Guangdong 518172, P. R. China}

\author{Zirui Sheng}
\affiliation{Guangdong Basic Research Center of Excellence for Aggregate Science, School of Science and Engineering, The Chinese University of Hong Kong (Shenzhen), Shenzhen, Guangdong 518172, P. R. China}

\author{Jianpeng Chen}
\affiliation{Guangdong Basic Research Center of Excellence for Aggregate Science, School of Science and Engineering, The Chinese University of Hong Kong (Shenzhen), Shenzhen, Guangdong 518172, P. R. China}

\author{Cunxi Gong}
\affiliation{Guangdong Basic Research Center of Excellence for Aggregate Science, School of Science and Engineering, The Chinese University of Hong Kong (Shenzhen), Shenzhen, Guangdong 518172, P. R. China}

\author{Weitang Li}
\email{liweitang@cuhk.edu.cn}
\affiliation{Guangdong Basic Research Center of Excellence for Aggregate Science, School of Science and Engineering, The Chinese University of Hong Kong (Shenzhen), Shenzhen, Guangdong 518172, P. R. China}


\begin{document}

\begin{abstract}
We present a scalable multi-GPU implementation of the time-dependent variational principle (TDVP) for multiset matrix product states (MPS) and tree tensor network states (TTNS) to simulate coupled electronic and vibrational dynamics. The multiset ans{\"a}tze offer a more accurate wavefunction representation for larger systems and longer propagation times, but their computational cost and memory requirements hinder large-scale applications. Our multi-GPU implementation parallelizes multiset TDVP across electronic states using batched tensor contractions and a distributed Krylov solver, while exploiting sparse electronic couplings to reduce inter-GPU communication. For a \(21\times21\) Holstein lattice, scaling from one to eight GPUs yields a \(4.5\times\) speedup and a \(7\times\) reduction in peak memory per GPU. With this new implementation, we are able to benchmark mixed quantum--classical method for large-scale nonadiabatic dynamics, assessing its accuracy and applicability across different electron--vibrational coupling regimes. Further numerical experiment on P3HT:PCBM heterojunction exciton dissociation and finite-temperature absorption and fluorescence spectra of aggregate show the accuracy–cost advantage and the versatility of our method.
\end{abstract}

\maketitle

\section{Introduction}
Accurate quantum dynamics is essential for understanding the mechanisms of many fundamental processes in molecular and condensed-matter systems, including exciton diffusion\cite{Sneyd2021,Giannini2022,Sheng2024FMO,Brey2024TransientLocalization}, charge transport\cite{Fratini2009,Fratini2017,Li2020CarrierMobility,Li2021charge,Takahashi2024HEOM,Jankovic2025}, spectroscopy\cite{Blanchet1999,Baiardi2013spectra,Ren2018Spectra,Yang2020Spectra,Barandun2026Anharmonic}, and open quantum systems\cite{Prior2010,Vega2018open,Strathearn2018,Tanimura2020,shi2022Open,Zhang2025msHEOM}. These processes often involve coupled electronic and vibrational degrees of freedom, posing the computational challenges due to the exponential growth of the underlying Hilbert space and the complexity of quantum correlations. Simulations of realistic molecular and material systems further require larger system sizes, longer propagation times, and finite-temperature treatments, placing additional demands on computational accuracy and efficiency.

Tensor network states, including matrix product states (MPS) and tree tensor network states (TTNS), provide compact representations of high-dimensional quantum wavefunctions through low-rank tensor decompositions\cite{Ostlund1995DMRG,Schollwock2005DMRG,PAECKEL2019167998,Ma2022DMRG,Xie2026TDDMRG}. For quantum dynamics involving coupled electronic states, the multiset ans{\"a}tze assign an independent vibrational tensor network state to each electronic state. These representations can achieve higher accuracy at a given bond dimension and extend accessible propagation times relative to singleset representation. Originally introduced within the multiconfiguration time-dependent Hartree (MCTDH) framework,\cite{BECK20001} the multiset ans{\"a}tze were later extended to MPS by Kloss et al. \cite{Kloss2019Multiset}, who employed the time-dependent variational principle (TDVP) for real time propagation. Subsequent developments encompassing TTNS and open quantum system further demonstrate their versatility.\cite{Lindoy2025pyTTN,Zhang2025msHEOM,lindoy2026adaptivemultisettreetensor} While their improved accuracy makes the multiset ans{\"a}tze particularly promising for large-scale quantum dynamics, storing the tensor network state and performing tensor contractions impose substantial memory and computational costs.

Graphics processing units (GPUs) are designed for massively parallel floating-point computations, making them well suitable for scientific computing applications. In \textit{ab initio} density matrix renormalization group (DMRG), advances in (multi-)GPU implementations have significantly improved the large-scale electronic structure calculations.\cite{Tian2022GPU,Xiang2024GPU,Brower2026GPU,Menczer2025GPU,Du2026VectorizedTensorNetworks} Together, these developments highlight the key challenges in scaling tensor-network calculations beyond a single GPU, where inefficient workload distribution, memory management, and inter-GPU communication can offset the benefits of parallel computation. For quantum dynamics simulation, even though GPU-accelerated TDVP has achieved widely recognized speedups,\cite{Li2020TDDMRGTimeEvolution,Lambertson2024GPUSpectra,Kessing2026paces,lindoy2026adaptivemultisettreetensor} scalable multi-GPU implementation remains underexplored. Recent advances have demonstrated the potential of parallelized quantum dynamics simulation,\cite{Secular2020ParallelTDVP,wang2025krylov,kapás2026diversifyingtimeevolutionmatrix} but leave open how to efficiently distribute wavefunctions across multiple GPUs without introducing additional approximations.

In this work, we develop a scalable multi-GPU multiset TDVP framework for MPS and TTNS within the Renormalizer package.\cite{Renormalizer,Ren2022renormalizer} The implementation exploits the multiset ans{\"a}tze to distribute tensor contractions and Krylov vectors across GPUs while minimizing inter-GPU communication, enabling accurate quantum dynamics simulations at an unprecedented scale. In addition, we develop algorithmic extensions of the multiset ans{\"a}tze for finite-temperature quantum dynamics and the calculation of absorption and fluorescence spectra. The rest of this paper is structured as follows. In Section~\ref{sec:theory}, we introduce the fundamentals of the multiset ans{\"a}tze and TDVP, their extensions to finite-temperature dynamics, and the multi-GPU parallelization strategies. In Section~\ref{sec:results}, we demonstrate the advantages of our approach in terms of scalability and computational efficiency through comprehensive benchmarks and illustrate its applications to chemical systems. Our large-scale fully quantum simulations reveal the high accuracy of mixed quantum--classical method in the weak electron--vibrational coupling regime, while highlighting the indispensable role of fully quantum dynamics under strong coupling. In Section~\ref{sec:conclusion}, we summarize the present methodology and provide an outlook on future developments and applications.

\section{Theory and Computational Methods}
\label{sec:theory}
\subsection{Singleset and Multiset MPS/TTNS Representations}
\label{sec:representations}

Matrix product states (MPS) and tree tensor network states (TTNS) represent high-dimensional wavefunctions as contractions of low-rank tensors. TTNS provides the more general tree-structured representation, whereas MPS is the special case in which the tensors form a one-dimensional chain. More detailed explanations can be found in several excellent literature.\cite{ren2022mpo,Hino2022MCTDH,Li2024TTNO,Larsson17072024,Dorfner2024Comparison,Xie2026TDDMRG,Li2026MpsTtns} For a quantum system with multiple physical degrees of freedom (DoFs), we consider a general tree tensor network consisting of \(N\) nodes. The corresponding TTNS wavefunction can be expressed as
\begin{equation}
    \ket{\Phi(t)}
    =
    \sum_{\{a\},\{\sigma\}}
    \left[\prod_{j=1}^{N} A[j]^{\sigma_j}_{\Lambda_j,a_j}(t)\right]
    \ket{\sigma_1\cdots\sigma_N}.
    \label{eq:vibrational-ttns}
\end{equation}
Here, \(\sigma_j\) labels the local basis for a DoF or a group of DoFs. For internal nodes that arise from the hierarchical grouping of lower-level DoFs and therefore do not correspond directly to a physical DoF, a trivial physical index of dimension one is assigned. With a root node chosen, \(a_j\) connects node \(j\) to its parent, while \(\Lambda_j\) denotes the set of virtual indices connecting it to its children. The root node is assigned a one-dimensional parent index, whereas \(\Lambda_j\) is empty for a leaf node. Summation over \(\{a\}\) then contracts the local tensors according to the tree topology.\cite{Larsson17072024} For a one-dimensional chain of \(L\) ordered sites, TTNS reduces to
\begin{equation}
    \ket{\Phi(t)}
    =
    \sum_{\{a\},\{\sigma\}}
    A[1]^{\sigma_1}_{a_0a_1}(t)
    A[2]^{\sigma_2}_{a_1a_2}(t)
    \cdots
    A[L]^{\sigma_L}_{a_{L-1}a_L}(t)
    \ket{\sigma_1\cdots\sigma_L},
    \label{eq:vibrational-mps}
\end{equation}
where \(a_0\) and \(a_L\) are one-dimensional boundary indices. In either representation, cutting a virtual bond separates the network into two parts, and the corresponding bond dimension \(\chi_j\) limits the bipartite entanglement that can be represented across the bond, thereby affecting both accuracy and computational cost. To fix the gauge redundancy, both MPS and TTNS can be transformed into canonical form with an orthogonality center at a chosen node, as discussed further in Sec.~\ref{sec:tdvp}.

On the same tree topology, an operator \(\hat O\) can be represented as a tree tensor network operator (TTNO),
\begin{equation}
    \hat O
    =
    \sum_{\{w\},\{\sigma\},\{\sigma'\}}
    \left[\prod_{j=1}^{N} W[j]^{\sigma_j'\sigma_j}_{\Omega_j,w_j}\right]
    \ket{\sigma_1'\cdots\sigma_N'}
    \bra{\sigma_1\cdots\sigma_N}.
    \label{eq:ttno-definition}
\end{equation}
Each local tensor \(W^{[j]}\) carries a pair of physical indices, \(\sigma_j'\) and \(\sigma_j\), associated with the local ket and bra basis states, together with virtual indices \(w_j\) and \(\Omega_j\) connecting the parent and child nodes, respectively. The dimension of \(w_j\) defines the operator bond dimension. For a one-dimensional chain, the TTNO reduces to a matrix product operator (MPO). Given a sum-of-product operator and a fixed network topology, optimal and exact MPO/TTNO representations can be constructed automatically.\cite{ren2022mpo,Li2024TTNO}

\begin{figure}[htbp]
    \centering
    \includegraphics[width=0.8\textwidth]{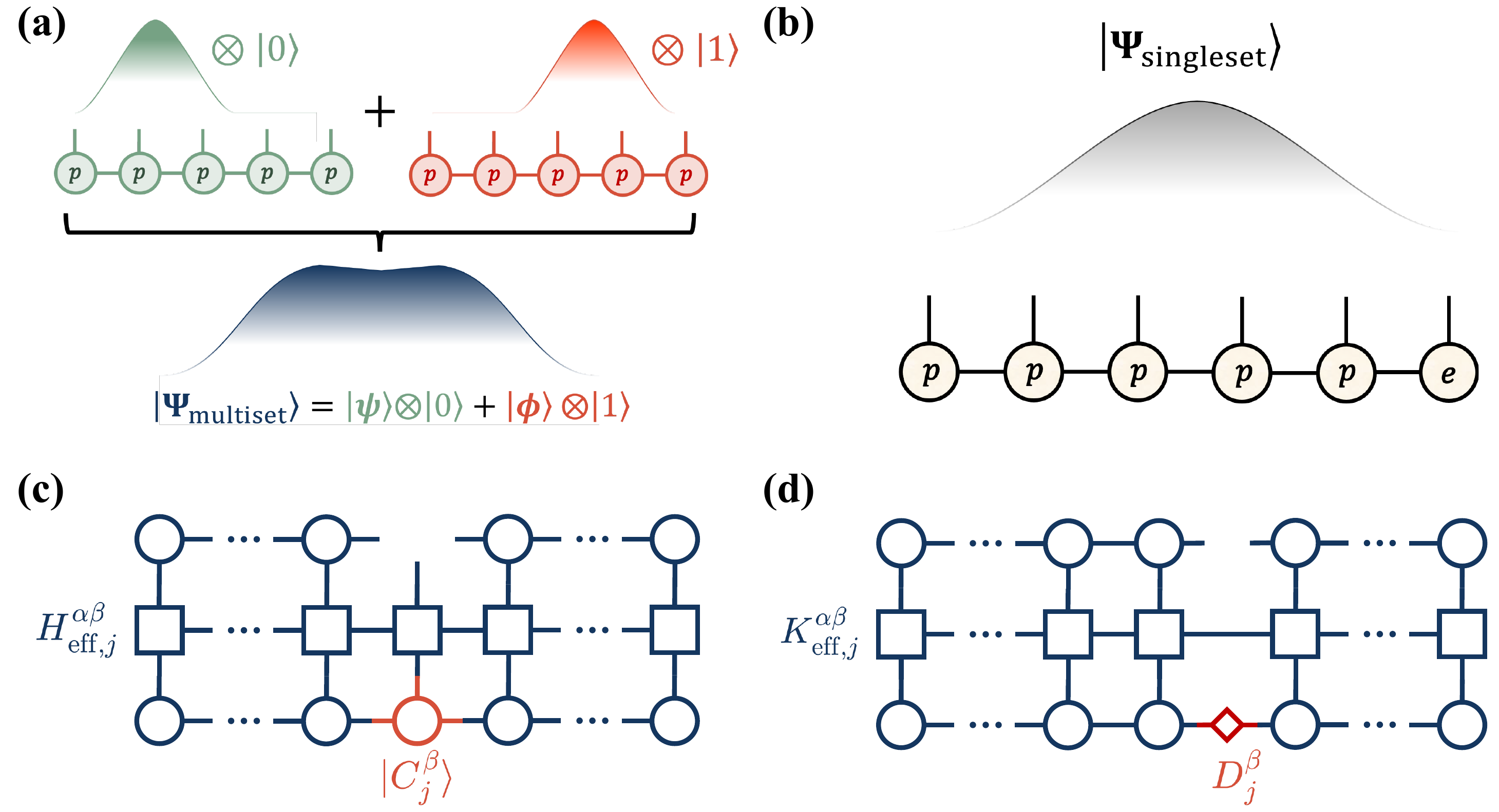}
    \caption{Schematic comparison of (a) multiset and (b) singleset MPS representations. The multiset ans{\"a}tze assign independent vibrational MPSs \(\ket{\psi}\) and \(\ket{\phi}\) for electronic states \(\ket{0}\) and \(\ket{1}\), allowing a more accurate representation than singleset MPS. Effective Hamiltonian contractions for (c) \(H_{\mathrm{eff},j}^{\alpha\beta}\ket{C_j^\beta}\) and (d) \(K_{\mathrm{eff},j}^{\alpha\beta}D_j^\beta\) in multiset TDVP.}
    \label{fig:multiset-mps}
    \label{fig:multiset-tdvp-projector}
\end{figure}

Considering a system with \(N_{\mathrm e}\) electronic basis states \(\{|\alpha\rangle\}\), the multiset ans{\"a}tze associate each basis state with an independently vibrational wavefunction,
\begin{equation}
    \ket{\Psi_{\mathrm{ms}}(t)}
    =
    \sum_{\alpha=1}^{N_{\mathrm e}}
    \ket{\alpha}\ket{\Phi^\alpha(t)}.
    \label{eq:multiset-ansatz}
\end{equation}
Each \(|\Phi^\alpha(t)\rangle\) is represented by a MPS or TTNS. Figure~\ref{fig:multiset-mps}(a,b) compares the multiset and singleset MPS representations for two electronic states. In the multiset representation, \(\ket{\psi}\) and \(\ket{\phi}\) independently describe the vibrational wavefunctions associated with electronic states \(\ket{0}\) and \(\ket{1}\), respectively. By contrast, the singleset ans{\"a}tze  incorporates both electronic and vibrational degrees of freedom into a single MPS, whose virtual bond spaces must accommodate the vibrational wavefunctions associated with different electronic states. Consequently, the multiset ans{\"a}tze can achieve higher accuracy with same bond dimensions when these vibrational wavefunctions develop distinct structures and correlations.

In the electronic basis, a general operator acting on the full Hilbert space can be written as
\begin{equation}
    \hat O
    =
    \sum_{\alpha,\beta=1}^{N_{\mathrm e}}
    \ket{\alpha}\!\bra{\beta}
    \otimes
    \hat O^{\alpha\beta},
    \label{eq:multiset-operator}
\end{equation}
where
\(\hat O^{\alpha\beta}=\bra{\alpha}\hat O\ket{\beta}\)
acts on the vibrational Hilbert space and can be represented as an MPO or TTNO. For the Hamiltonian, the component of \(\hat H\ket{\Psi_{\mathrm{ms}}(t)}\) associated with electronic state \(\alpha\) is \(\sum_{\beta}\hat H^{\alpha\beta}\ket{\Phi^\beta(t)},\)
which couples the independent vibrational wavefunctions associated with different electronic states.

\subsection{Multiset Projector-Splitting TDVP}
\label{sec:tdvp}

We propagate the multiset MPS/TTNS using the one-site projector-splitting TDVP integrator. TDVP projects the wavefunction time derivative closest to the exact Schr\"odinger evolution within the tangent space of the chosen tensor-network manifold,\cite{Haegeman2016TDVP,Kloss2019Multiset,Lindoy2025pyTTN}
\begin{equation}
    i\frac{\partial}{\partial t}\ket{\Psi_{\mathrm{ms}}(t)}
    =\hat P_{T_{\Psi_{\mathrm{ms}}}\mathcal M_{\mathrm{ms}}}
    \hat H\ket{\Psi_{\mathrm{ms}}(t)},
    \label{eq:tdvp-projected-schrodinger}
\end{equation}
In multiset MPS \(\ket{\Psi_{\mathrm{ms}}(t)}\), each vibrational MPS \(\ket{\Phi^\alpha(t)}\) can be written in mixed-canonical form at site \(j\) as
\(
\ket{\Phi^\alpha(t)}
=
\ket{L_j^\alpha}
\ket{C_j^\alpha(t)}
\ket{R_j^\alpha}.
\) Here, \(C_j^\alpha\) is the center of orthogonality at site \(j\). The left-canonical tensors on sites \(1,\ldots,j-1\) and the right-canonical tensors on sites \(j+1,\ldots,L\) define the \(\ket{L_j^\alpha}\) and \(\ket{R_j^\alpha}\), respectively. In these bases, the multiset tangent-space projector takes the form

\begin{equation}
\begin{aligned}
    \hat P_{T_{\Psi_{\mathrm{ms}}}\mathcal M_{\mathrm{ms}}}
    =\sum_{\alpha=1}^{N_{\mathrm e}}\ket{\alpha}\bra{\alpha}\otimes
    \Bigg[&\sum_{j=1}^{L}
    \ket{L_j^\alpha}\bra{L_j^\alpha}\otimes\hat I_j\otimes
    \ket{R_j^\alpha}\bra{R_j^\alpha}\\
    &-\sum_{j=1}^{L-1}
    \ket{L_{j+1}^\alpha}\bra{L_{j+1}^\alpha}\otimes
    \ket{R_j^\alpha}\bra{R_j^\alpha}\Bigg],
\end{aligned}
    \label{eq:multiset-tangent-projector}
\end{equation}
Acting the projector in Eq.~\eqref{eq:multiset-tangent-projector} on \(\hat H\ket{\Psi_{\mathrm{ms}}}\) separates the projected Hamiltonian into two contributions. For the first term, the corresponding effective Hamiltonian can be defined as \(
H_{\mathrm{eff},j}^{\alpha\beta}
=
\bra{L_j^\alpha}\bra{R_j^\alpha}
\hat H^{\alpha\beta}
\ket{L_j^\beta}\ket{R_j^\beta}.
\) After shifting the orthogonality center to bond \((j,j+1)\), the second term gives \(
K_{\mathrm{eff},j}^{\alpha\beta}
=
\bra{L_{j+1}^\alpha}\bra{R_j^\alpha}
\hat H^{\alpha\beta}
\ket{L_{j+1}^\beta}\ket{R_j^\beta}.
\)
Figure~\ref{fig:multiset-tdvp-projector}(c,d) illustrates the action of these effective Hamiltonians on the site and bond centers.

A symmetric splitting of the two contributions in Eq.~\eqref{eq:multiset-tangent-projector} yields a second-order projector-splitting integrator over a time step \(\tau\). During a left-to-right half sweep, the center tensor \(C_j^\alpha\) is first propagated forward by \(\tau/2\) according to
\begin{equation}
    i\ket{\dot C_j^\alpha(t)}
    =\sum_{\beta=1}^{N_{\mathrm e}}H_{\mathrm{eff},j}^{\alpha\beta}\ket{C_j^\beta(t)}.
    \label{eq:multiset-site-eom}
\end{equation}
A QR factorization of the updated \(C_j^\alpha\) produces a left-canonical tensor and the bond matrix \(D_j^\alpha\), placing the wavefunction in the form \(\ket{\Phi^\alpha(t)}=\ket{L_{j+1}^\alpha}D_j^\alpha(t)\ket{R_j^\alpha}\). Since the second part in Eq.~\eqref{eq:multiset-tangent-projector} enters with a minus sign, \(D_j^\alpha\) is then propagated backward by \(\tau/2\) according to
\begin{equation}
    i\dot D_j^\alpha(t)
    =\sum_{\beta=1}^{N_{\mathrm e}}K_{\mathrm{eff},j}^{\alpha\beta}D_j^\beta(t).
    \label{eq:multiset-bond-eom}
\end{equation}
The propagated \(D_j^\alpha(t)\) is contracted into site \(j+1\), and the same sequence is repeated as the orthogonality center advances along the chain. Algorithm~S2 in the Supporting Information gives the complete sequence. The sum over \(\beta\) couples the equations for all electronic states, which are propagated jointly using a Lanczos--Krylov solver.

For TTNS, the left and right chain bases are replaced by the renormalized bases of the subtrees attached to the active node. Site and bond updates become node and edge updates along the tree, while the electronic state coupling is unchanged.\cite{Lindoy2025pyTTN,Li2026MpsTtns} The initial bond expansion used before real time propagation is described in the Supporting Information.

\subsection{Finite-Temperature Purification and Dipole Correlation Functions}
\label{sec:tfd}
Finite-temperature quantum simulations have been developed using various numerical approaches. Within tensor-network frameworks, thermal density operators can be represented as matrix product operator, with the thermal states constructed through imaginary-time evolution.\cite{Verstraete2004MPDO,Feiguin2005Purification,Li2023tanTRG,Zhang2026EnhancedXTRG,Li2026FixedNtanTRG} For coupled electron–vibrational systems, thermo-field dynamics (TFD) provides a more economical realization, which incorporates the temperature parameter through the Bogoliubov transformation in an enlarged physical--auxiliary bosonic space.\cite{Borrelli2016TFD,Borrelli2021TFDReview,Zhang2024CoherenceTFD}

At finite temperature, thermal equilibrium is described by the mixed-state density operator \(\hat{\rho}_\beta=e^{-\beta\hat H}/Z\), with \(\beta=(k_{\mathrm B}T)^{-1}\), rather than by a single wavefunction in the physical Hilbert space \(P\). A pure-state representation can be introduced by enlarging the Hilbert space with an auxiliary space \(Q\),
\begin{equation}  
    \hat{\rho}_\beta
    =
    \operatorname{Tr}_{Q}
    \ket{\Psi_\beta}\bra{\Psi_\beta}.
\end{equation}

One standard construction of a purified thermal state starts from the infinite-temperature limit. In the enlarged Hilbert space, the corresponding reference is the unnormalized maximally entangled state \(\ket{I}_{PQ}=\sum_n\ket{n}_P\ket{n}_Q\). Imaginary-time evolution of the physical degrees of freedom over \(\beta/2\) then gives the finite-temperature purified state,

\begin{equation}
    \ket{\Psi_\beta}_{PQ}
    \propto
    e^{-\beta\hat H/2}
    \ket{I}_{PQ},
    \qquad
    \beta=(k_{\mathrm B}T)^{-1},
\end{equation}
where \(k_{\mathrm B}\) is the Boltzmann constant. The proportionality reflects the normalization of the purified state, so the partition function need not be evaluated explicitly. Once prepared, \(\ket{\Psi_\beta}\) serves as the finite-temperature reference state for subsequent real-time propagation.

In the multiset representation of the purified state, the physical and auxiliary electronic indices are resolved explicitly,
\begin{equation}
   \ket{\Psi_\beta}=
   \sum_{\alpha,\eta=1}^{N_{\mathrm e}}
   \ket{\alpha}_P
   \ket{\eta}_Q
   \ket{\Phi_\beta^{\alpha\eta}}{\mathrm{vib}},
   \label{eq:thermal-multiset-state}
\end{equation}
where \(\eta\) labels the auxiliary electronic state, and \(\ket{\Phi_\beta^{\alpha\eta}}_{\mathrm{vib}}\) is represented by
\(N_{\mathrm e}^2\) independent vibrational MPS/TTNSs. The Hamiltonian acts exclusively in the physical space \(P\), so that \(\hat H^{\alpha\beta}\) connect only the physical electronic states, while the auxiliary space \(Q\) does not participate in the real time evolution. Consequently, observables are obtained by tracing the purified state over the auxiliary degrees of freedom, with the electronic part of this trace corresponding to a sum over \(\eta\).

For systems in which thermal excitation of electronic states is negligible and temperature effects are predominantly influences the vibrational degrees of freedom, thermo-field dynamics (TFD) provides a more economical purification scheme. In this formulation, the thermal vibrational state is represented by a temperature-dependent vacuum in an enlarged physical--auxiliary vibrational space. For a set of harmonic modes, this thermal vacuum can be written as
\begin{equation}
    \ket{0(\beta)}
    =
    \prod_k
    \frac{1}{\sqrt{Z_k}}
    \sum_{n=0}^{d_k-1}
    e^{-\beta\omega_k n/2}
    \ket{n_k}_P
    \ket{\widetilde n_k}_Q ,
\end{equation}
where \(d_k\) is the retained local basis dimension of mode \(k\). Equivalently, the same state can be generated from the vacuum of the doubled vibrational space through a Bogoliubov thermal transformation, 
\begin{equation}
    \ket{0(\beta)}
    =
    e^{-i\hat G}\ket{0,\widetilde 0},
    \qquad
    \hat G
    =
    -i\sum_k
    \theta_k
    \left(
    b_k\widetilde b_k
    -
    b_k^\dagger\widetilde b_k^\dagger
    \right),
\end{equation}
where \(\theta_k=\operatorname{arctanh}\left(e^{-\beta\omega_k/2}\right).\) The temperature dependence can therefore be transferred from the initial thermal vacuum to the vibrational operators, yielding a transformed Hamiltonian,
\begin{equation}  
    \hat H_\theta^{\alpha\beta}
    =
    e^{i\hat G}
    \left(
    \hat H^{\alpha\beta}
    -
    \delta_{\alpha\beta}\hat{\tilde H}_{\mathrm{vib}}
    \right)
    e^{-i\hat G},
\end{equation}
where \(\hat{\tilde H}_{\mathrm{vib}}\) is the auxiliary vibrational part of \(\hat H^{\alpha\beta}\) and \(\delta_{\alpha\beta}\) is the Kronecker delta. Under this condition, Borrelli et al. \cite{Borrelli2016TFD} showed that the auxiliary electronic degrees of freedom can be eliminated, which reduces the finite-temperature representation from \(N_{\mathrm e}^{2}\) to \(N_{\mathrm e}\) vibrational MPS/TTNSs.

The absorption and fluorescence spectra follow from finite temperature dipole correlation functions.\cite{Ren2018Spectra,Barandun2026Anharmonic} Let \(\hat H_g\) and \(\hat H_e\) be the zero-exciton vibrational and full one-exciton Hamiltonians, respectively. Taking fluorescence spectra as an example, the thermal reduced density matrix is \(\hat\rho_e^\beta=e^{-\beta\hat H_e}/Z_e\). With \(\hat\mu_{eg}\) mapping \(g\) to \(e\) and \(\hat\mu_{ge}=\hat\mu_{eg}^\dagger\), the correlation function with respect to \(|\alpha\rangle\) and \(|\beta\rangle\) is
\begin{equation}
    C^{\alpha\beta}(t)
    =\operatorname{Tr}_e\!\left[
    \hat\rho_e^\beta
    e^{i\hat H_e t/\hbar}
    \hat\mu_{\alpha g}
    e^{-i\hat H_g t/\hbar}
    \hat\mu_{g\beta}
    \right],
    \label{eq:thermal-fluorescence-correlation}
\end{equation}
where \(\hat{\mu}_{\alpha g}=|\alpha\rangle\langle\alpha|\hat{\mu}_{eg}\), and \(\mathrm{Tr}_e\) denotes the trace over the one-exciton Hilbert space. Subsequently, the corresponding spectral contributions are obtained by Fourier transforming the correlation functions,
\begin{equation}
    I^{\alpha\beta}(E)
    \propto
    \operatorname{Re}
    \int_0^\infty
    dt\,
    e^{-iEt/\hbar}
    C^{\alpha\beta}(t),
    \label{eq:spectral-line-strength}
\end{equation}
 With \(E=\hbar\omega\), the total fluorescence intensity is obtained by summing over \(\alpha\) and \(\beta\) with the frequency factor \(\omega^3\), yielding \(I_{\mathrm{fl}}(\omega)\propto\omega^3\sum_{\alpha,\beta}I^{\alpha\beta}(\hbar\omega)\). The absorption spectrum is evaluated in a similar manner, but with the ground state thermal reference \(\hat{\rho}_g^\beta=e^{-\beta\hat{H}_g}/Z_g\) and a frequency factor of \(\omega\).

\subsection{Scalable Multi-GPU Design for Multiset TDVP}
\label{sec:gpu}

\begin{figure}[htbp]
    \centering
    \includegraphics[width=1\textwidth]{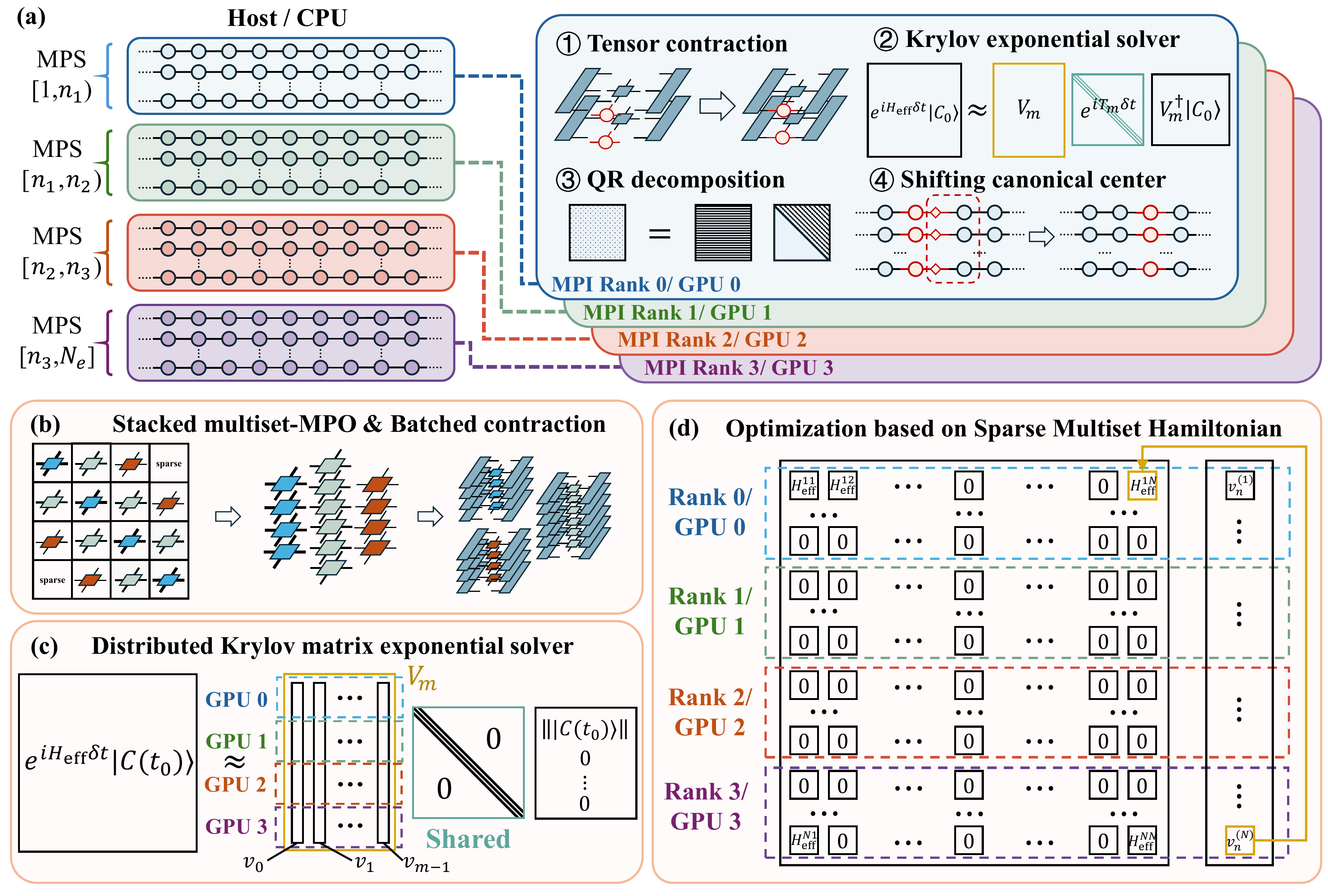}
    \caption{Schematic of the multi-GPU implementation of multiset TDVP. (a) MPSs are distributed across four MPI ranks, each assigned to one GPU. Long-lived tensors are stored in host memory, while tensors required for the current TDVP update are transferred to the GPU. (b) MPO tensors and contraction operands with matching dimensions are grouped for batched contractions on each GPU. (c) The Krylov basis \(V_m\) is distributed across GPUs, while the projected matrix \(T_m\) and normalization quantities are shared during Lanczos iterations. (d) For sparse multiset Hamiltonians, only the Krylov vectors required by nonzero electronic couplings between ranks are communicated, as highlighted by the yellow boxes and arrow. Here, \(v_n^{(1)}\) denotes the component of the \(n\)th Krylov vector corresponding to the first vibrational MPS.}
    \label{fig:multiGPU_workflow}
\end{figure}

Extending multiset TDVP to larger systems is limited by both memory capacity and computational throughput. Increasing the physical system simultaneously increases the number of MPSs, MPOs, and Krylov vectors, so the tensor workload can exceed the memory available to a single accelerator. Moreover, efficient GPU acceleration relies on regular batched operations over tensors with identical shapes. In multiset TDVP, however, tensor dimensions vary with the sweep position, complicating the use of uniform batched operations. Our implementation employs the Message Passing Interface (MPI) to organize memory and computation across multiple GPUs, as illustrated in Fig.~\ref{fig:multiGPU_workflow}(a). Each MPI rank is paired with one GPU and owns a set of the MPSs. Specifically, long-lived tensors are retained in the host memory of each rank, with only the tensors required at the current sweep position are staged on the GPU. All ranks advance the same MPS site and remain synchronized throughout the left-to-right and right-to-left TDVP sweeps. In this way, the multiset ans{\"a}tze distributes MPSs across GPUs while preserving the sequential structure of the TDVP integrator. Within this framework, the remaining computational challenges are addressed by batched MPO contractions, distributed Krylov matrix exponential solver, and the optimization based on sparsity, as detailed in Secs.~\ref{sec:gpu-batching}--\ref{sec:gpu-sparse}, respectively.

\subsubsection{Batched MPO Contractions}
\label{sec:gpu-batching}

The batch MPO contraction in Fig.~\ref{fig:multiGPU_workflow}(b) addresses the heterogeneity of tensor dimensions in the local multiset operators. Although the Hamiltonian contains \(N_{\mathrm e}\times N_{\mathrm e}\) terms in the multiset ans{\"a}tze, the corresponding MPO tensors generally fall into a small number of recurring shapes. At each sweep position, MPO tensors with matching dimensions are grouped and stacked, together with their corresponding MPS tensors and left and right environments arranged in the same order. Each group is then evaluated using batched contractions, reducing the overhead of repeatedly scheduling individual tensor contractions while improving GPU utilization.

After propagation, updated canonical-center tensors with compatible matrix dimensions are likewise factorized using batched QR. During the first sweep, the implementation records the grouping pattern together with the workspace layout and buffer sizes required at each sweep position. These buffers can then be preallocated and reused in subsequent sweeps. Shape grouping therefore makes the tensor workload on each GPU more regular, while workspace reuse reduces repeated memory allocation. These optimizations improve computational efficiency without modifying the projected TDVP equations.

\subsubsection{Distributed Krylov Matrix Exponential Solver}
\label{sec:gpu-krylov}

The multiset TDVP equations require repeated matrix exponential actions, which are evaluated using a Krylov solver. At a given multiset TDVP substep, the canonical centers of all MPSs are stacked into a single vector \(C(t_0)\), whose normalized form defines the initial Krylov vector \(v_0\). Repeated applications of \(H_{\mathrm{eff}}\) generate the Krylov subspace
\begin{equation}
    \mathcal V_n
    =
    \operatorname{span}
    \left\{
        C(t_0),
        H_{\mathrm{eff}}C(t_0),
        \ldots,
        H_{\mathrm{eff}}^{n-1}C(t_0)
    \right\}.
    \label{eq:distributed-krylov-space}
\end{equation}
Lanczos orthogonalization gives \(V_n=[v_1,\ldots,v_{n}]\) and the projected matrix \(T_n=V_n^\dagger H_{\mathrm{eff}}V_n\). The basis is enlarged iteratively until the projected Krylov representation converges to the exponential action \(e^{-iH_{\mathrm{eff}}\delta t}C(t_0)\).

Besides the repeated tensor contractions required for \(H_{\mathrm{eff}}v_n\), constructing and storing the Krylov basis constitutes another major GPU workload. Since the basis is generated recursively, with each new vector depending on the preceding ones, straightforward multi-GPU parallelization is difficult, limiting both computational efficiency and accessible GPU memory. Figure~\ref{fig:multiGPU_workflow}(c) illustrates our distributed Krylov procedure. Each GPU stores only its portion of \(V_n\) and evaluates the corresponding part of \(H_{\mathrm{eff}}v_n\). Contributions from all GPUs are combined to obtain the normalization scalars, which is then shared to normalize each newly generated Krylov vector. Since these scalars and \(T_n\) are much smaller than \(V_n\), they can be shared at low communication cost while the Krylov basis remains distributed. Detailed pseudocode for this distributed Lanczos--Krylov procedure is provided as Algorithm~S1 in the Supporting Information.

\subsubsection{Optimization based on Sparse Multiset Hamiltonian}
\label{sec:gpu-sparse}

In typical molecular, aggregate, and lattice models, only a part of off-diagonal electronic couplings are nonzero, giving rise to a sparse multiset Hamiltonian. This sparsity is particularly important in the Krylov solver, where the tensor contractions required for \(H_{\mathrm{eff}}v_n\) are repeated at every Krylov iteration. Retaining only the nonzero \(H^{\alpha\beta}\) terms reduces unnecessary tensor operations and identifies the data dependencies required for inter-GPU communication. 

Accordingly, contractions associated with vanishing \(H^{\alpha\beta}\) terms are omitted. As illustrated in Fig.~\ref{fig:multiGPU_workflow}(d), each rank computes a portion of \(H_{\mathrm{eff}}v_n\) corresponding to its assigned electronic states, with the Krylov vectors partitioned into components from \(v_n^{(1)}\) to \(v_n^{(N)}\). When a nonzero \(H^{\alpha\beta}\) couples electronic states assigned to different ranks, only the required parts of \(v_n\) are communicated. For example, as illustrated by the yellow boxes and arrow in Fig.~\ref{fig:multiGPU_workflow}(d), the contracted tensor \(H_{\mathrm{eff}}^{1N}v_n^{(N)}\) remains on rank 0 for subsequent computations, while the Hamiltonian MPOs and environments remain local to the ranks performing the contractions. Unlike approaches that replicate the entire Krylov basis across ranks, our optimization based on sparse multiset Hamiltonian eliminates the need to synchronize full Krylov vectors across GPUs, thereby avoiding additional inter-GPU communication of large tensors.

\section{Results and Discussion}
\label{sec:results}

\subsection{Scalability and Performance of Multi-GPU Multiset Ansatz}
\label{sec:gpu-performance}

In this section, we benchmark the scalability and computational performance of the multi-GPU multiset TDVP framework using the Frenkel–Holstein model. The model describes an electronic excitation moving between coupled sites and interacting with local harmonic vibrations. Within the single-exciton manifold, the Hamiltonian is given by
\begin{align}
    \hat H
    ={}&
    \sum_{\alpha=1}^{N_{\mathrm e}}
    \varepsilon_\alpha
    \hat a_\alpha^\dagger \hat a_\alpha
    +
    \sum_{\substack{\alpha,\beta=1\\\alpha\neq\beta}}^{N_{\mathrm e}}
    J_{\alpha\beta}
    \hat a_\alpha^\dagger \hat a_\beta
    +
    \sum_{\alpha k}
    \omega_{\alpha k}
    \hat b_{\alpha k}^\dagger \hat b_{\alpha k}
    \nonumber\\
    &+
    \sum_{\alpha k}
    \omega_{\alpha k}g_{\alpha k}
    \hat a_\alpha^\dagger \hat a_\alpha
    \left(
        \hat b_{\alpha k}^\dagger+\hat b_{\alpha k}
    \right).
    \label{eq:frenkel-holstein-hamiltonian}
\end{align}
where \(\hat a_\alpha^\dagger\) (\(\hat a_\alpha\)) creates (annihilates) a local Frenkel excitation at site \(\alpha\), and \(\hat b_{\alpha k}^\dagger\) (\(\hat b_{\alpha k}\)) creates (annihilates) a quantum of the \(k\)th intramolecular vibrational mode locally coupled to that excitation. The electronic part is specified by the local excitation energies \(\varepsilon_\alpha\) and the excitonic couplings \(J_{\alpha\beta}\) between sites, while \(\omega_{\alpha k}\) and \(g_{\alpha k}\) give the frequency and dimensionless coupling strength of the \(k\)th vibrational mode at site \(\alpha\), respectively. Electronic transfer promotes delocalization, while electron–vibrational coupling induces local nuclear reorganization with a reorganization energy \(\lambda_\alpha=\sum_k\omega_{\alpha k}g_{\alpha k}^2\). This interplay makes the model a useful setting for investigating how nuclear motion controls excitation transport and relaxation in molecular aggregates and related condensed-phase systems.

A growing body of recent work has established that dimensionality play a central role in microscopic mechanisms of molecular systems,\cite{Li2021charge} making two-dimensional lattices a natural step beyond one-dimensional model systems. We therefore consider two-dimensional square lattices with open boundary conditions and uniform nearest-neighbor excitonic couplings. For an \(L \times L\) lattice, row-by-row ordering places vertically adjacent sites \(L\) positions apart along the MPS chain. Consequently, entanglement between these sites must be encoded across the intervening MPS virtual bonds, increasing the bond dimensions needed to represent spatially extended correlations during time evolution.\cite{Stoudenmire2012,Banuls2023} These features make the two-dimensional Holstein model a demanding benchmark for large-scale quantum dynamics simulations.

\begin{figure}[htbp]
    \centering
    \includegraphics[width=\textwidth]{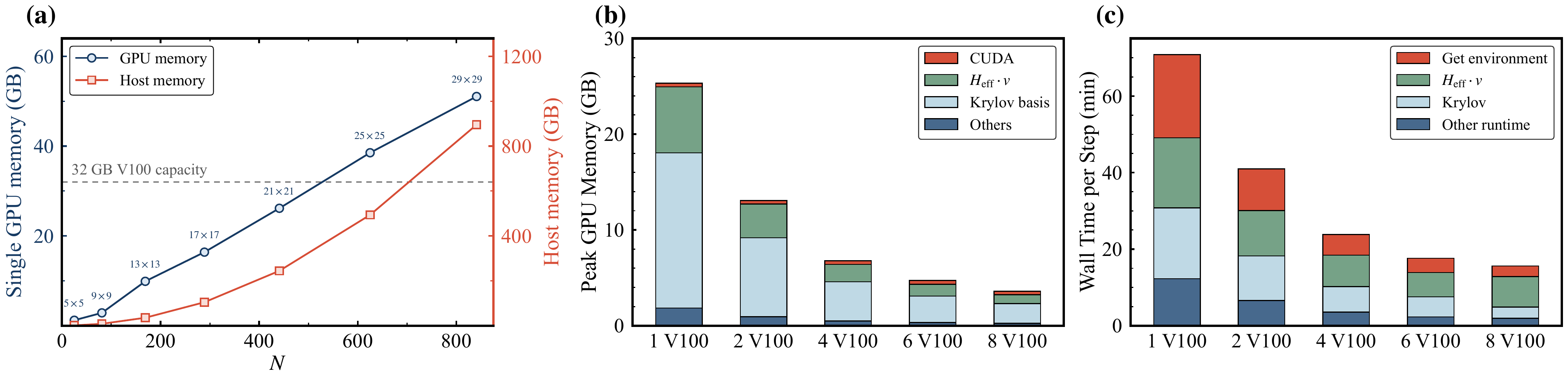}
    \caption{Memory scaling and multi-GPU performance of multiset TDVP. (a) Single-GPU and host memory test on NVIDIA A100 (80GB) for two-dimensional Holstein lattices. The horizontal line denotes the \(32\)~GiB nominal capacity of a NVIDIA V100. (b) Breakdown of peak memory per GPU and (c) breakdown of wall time per TDVP step for the \(21\times21\) lattice with \(\chi=64\) on one to eight NVIDIA V100 GPUs.}
    \label{fig:gpu-performance}
\end{figure}

To assess memory scaling with system size, we examine the GPU and host memory requirements for two-dimensional Holstein lattices, as shown in Figure~\ref{fig:gpu-performance}(a). For these calculations, we set \(J=\omega_0=1\), \(g=0.5\), and retain eight Fock states per vibrational mode, using MPS topology with \(\chi=64\). We initialize the Krylov subspace with 24 Lanczos vectors and add another 24 vectors if successive approximations fail to meet the relative tolerance of \(10^{-8}\) or absolute tolerance of \(10^{-10}\). Under these settings, the blue curve in Figure~\ref{fig:gpu-performance}(a) shows that the  single-GPU memory requirement grows linearly with system size, due to the Krylov basis and intermediate tensors for \(H_{\mathrm{eff}}v\) contractions dominating the GPU memory. Specifically, the memory requirement increases from \(25.3\)~GB for the \(21\times21\) lattice to \(36.6\)~GB for the \(25\times25\) lattice, exceeding the 32 GB memory capacity of a single NVIDIA V100 GPU. 

The host memory, in turn, retains the MPS tensors and environment tensors for all multiset Hamiltonian terms. These environments are updated incrementally during TDVP sweeps to avoid repeated tensor contractions, at the expense of additional memory storage. For nearest-neighbor electronic couplings, the number of nonzero Hamiltonian terms scales as \(O(N)\), with each term assigning \(O(N)\) environment tensors along the MPS chain, resulting in quadratic growth in host memory requirements. We acknowledge that the current host memory management strategy is not fully optimized and could be further improved through out-of-core tensor storage using GPUDirect Storage,\cite{NvidiaGDSOverview,Sheng2023FlexGen} which allows tensors to be transferred directly from NVMe storage to GPU memory without intermediate staging in host memory. A very recent study introduced an environment checkpointing scheme that retains only selected contraction environments and reconstructs the remaining ones on demand,\cite{Shrivastava2026EnvCheck} suggesting a general route to mitigating the host-memory growth. In the present work, we address single-GPU memory limitations through multi-GPU parallelization, thereby extending the accessible system sizes.

As depicted in Figure~\ref{fig:gpu-performance}(b) and (c), we benchmark a \(21\times21\) Holstein lattice that fits within the memory of a single NVIDIA V100 GPU, allowing direct comparisons using one to eight GPUs. For each GPU configuration, we monitor the GPU memory usage over the first five TDVP steps and report the maximum peak memory across all MPI ranks. The mean wall time per TDVP step is obtained by averaging the execution times over the same tasks. For a closer inspection of GPU memory usage, we report a detailed breakdown of peak device memory. Distributing the Krylov basis and intermediate tensors required for \(H_{\mathrm{eff}}v\) contractions across multiple GPUs substantially reduces the peak device memory, achieving a 7 times reduction from \(25.3\) GB on one V100 to \(3.6\) GB on eight NVIDIA V100 GPUs. The CUDA remains nearly constant at \(0.38\) GiB per GPU, primarily reflecting GPU context management during MPI process. In addition, the other memory includes memory allocated for batched operation buffers, QR update, and other small tensors. 

Besides memory consumption, multi-GPU parallelization accelerates the dominant TDVP operations, including effective Hamiltonian contractions, Krylov propagation, and environment updates. Other runtime includes QR factorization, tensor packing, normalization, and observable evaluation. For the present benchmark, four GPUs provide a favorable balance between computational speedup and parallel efficiency. Beyond four GPUs, the time attributed to \(H_{\mathrm{eff}}v\) contractions decreases only marginally and even increases slightly from six to eight GPUs. This behavior arises from the inter-rank communication described in Section~\ref{sec:gpu-sparse}, which becomes increasingly significant at larger MPI rank counts and limits linear scaling, despite our optimization that avoids synchronization of the full Krylov vectors. Ultimately, the mean wall time per TDVP step decreases from \(70.8\) min on one GPU to \(23.9\) and \(15.6\) min on four and eight GPUs, respectively, corresponding to speedups of \(3\) and \(4.5\).

\begin{figure}[htbp]
    \centering
    \includegraphics[width=0.4\textwidth]{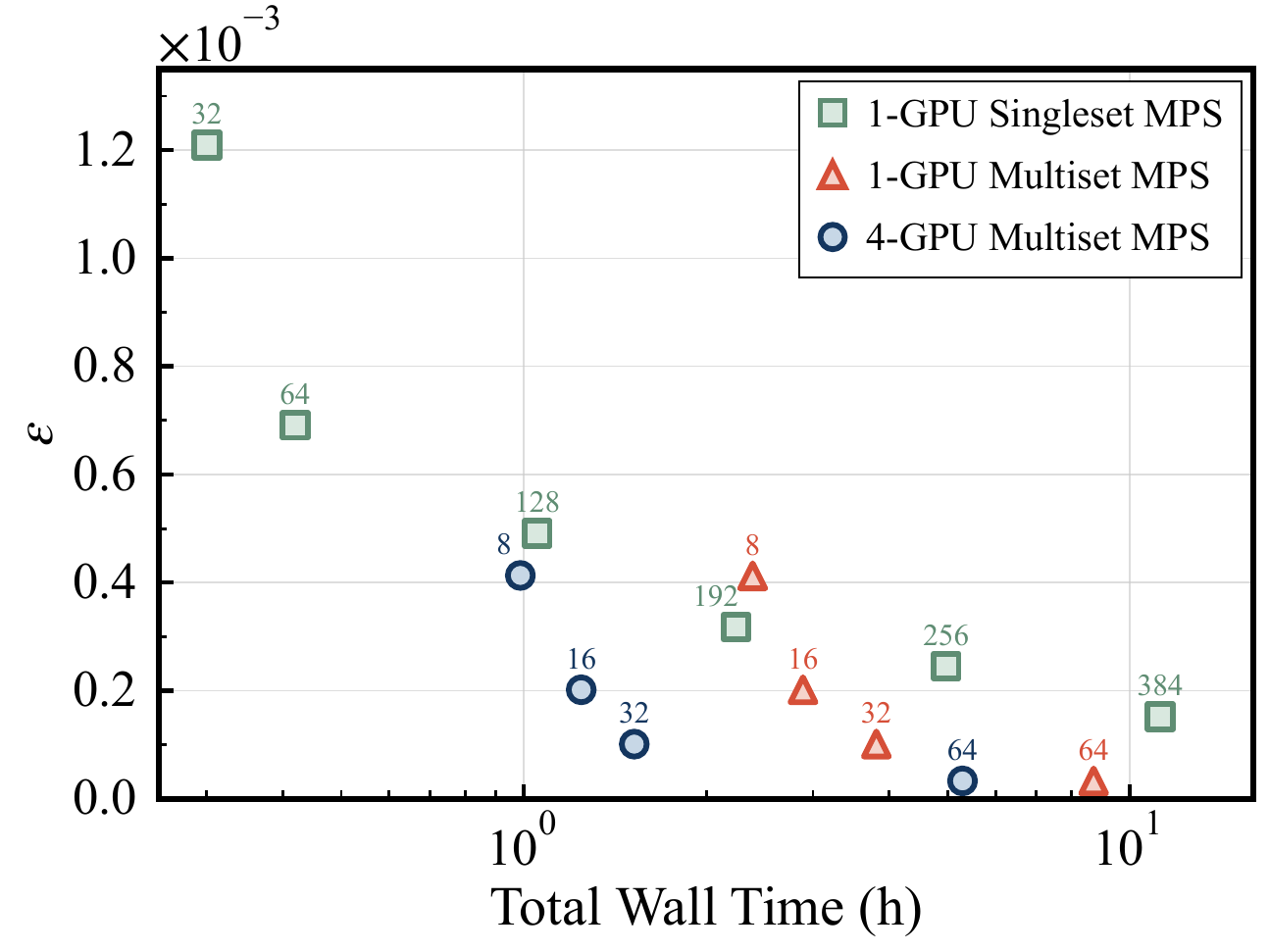}
    \caption{Accuracy versus propagation wall time for singleset and multiset MPS on the \(11\times11\) Holstein lattice. Each marker is annotated with the bond dimension \(\chi\).}
    \label{fig:holstein-accuracy}
\end{figure}

We next assess the performance of the multi-GPU implementation on the accuracy--efficiency trade-off for a separate 2D \(11\times11\) lattice with \(J=\omega_0=1\), \(g=1\), 16 Fock states per mode, and a time step of \(\Delta t=0.1\). We define the mean absolute population deviation from a reference standard as
\begin{equation}
    \epsilon(T) = \frac{1}{N_{\mathrm e}T}
    \int_0^T \sum_{i=1}^{N_{\mathrm e}}
    \left|\rho_i(t)-\rho_i^{\mathrm{ref}}(t)\right|\,\mathrm{d}t .
    \label{eq:cumulative-population-error}
\end{equation}
Here, \(N_{\mathrm e}\) is the number of electronic states. For Figure~\ref{fig:holstein-accuracy}, we use the four-GPU multiset MPS with \(\chi=128\) as the reference. The wall time includes the first 100 TDVP steps to \(T=10\). At a fixed bond dimension of \(\chi=32\), increasing the GPU count from one to four reduces the total wall time from \(3.8\) to \(1.5\)~h, while \(\epsilon\) remains close to \(1\times10^{-4}\). For comparison, the one-GPU singleset MPS with \(\chi=384\) requires \(11.2\)~h and gives \(\epsilon=1.5\times10^{-4}\). This indicates that our multi-GPU multiset implementation achieves higher accuracy at lower computational cost than singleset MPS, demonstrating a more favorable accuracy–cost trade-off.

\subsection{Large-Scale Exciton Dynamics in Two-Dimensional Holstein Lattices}
\label{sec:holstein-large-scale}

\begin{figure}[htbp]
    \centering
    \includegraphics[width=\textwidth,height=0.77\textheight,keepaspectratio]{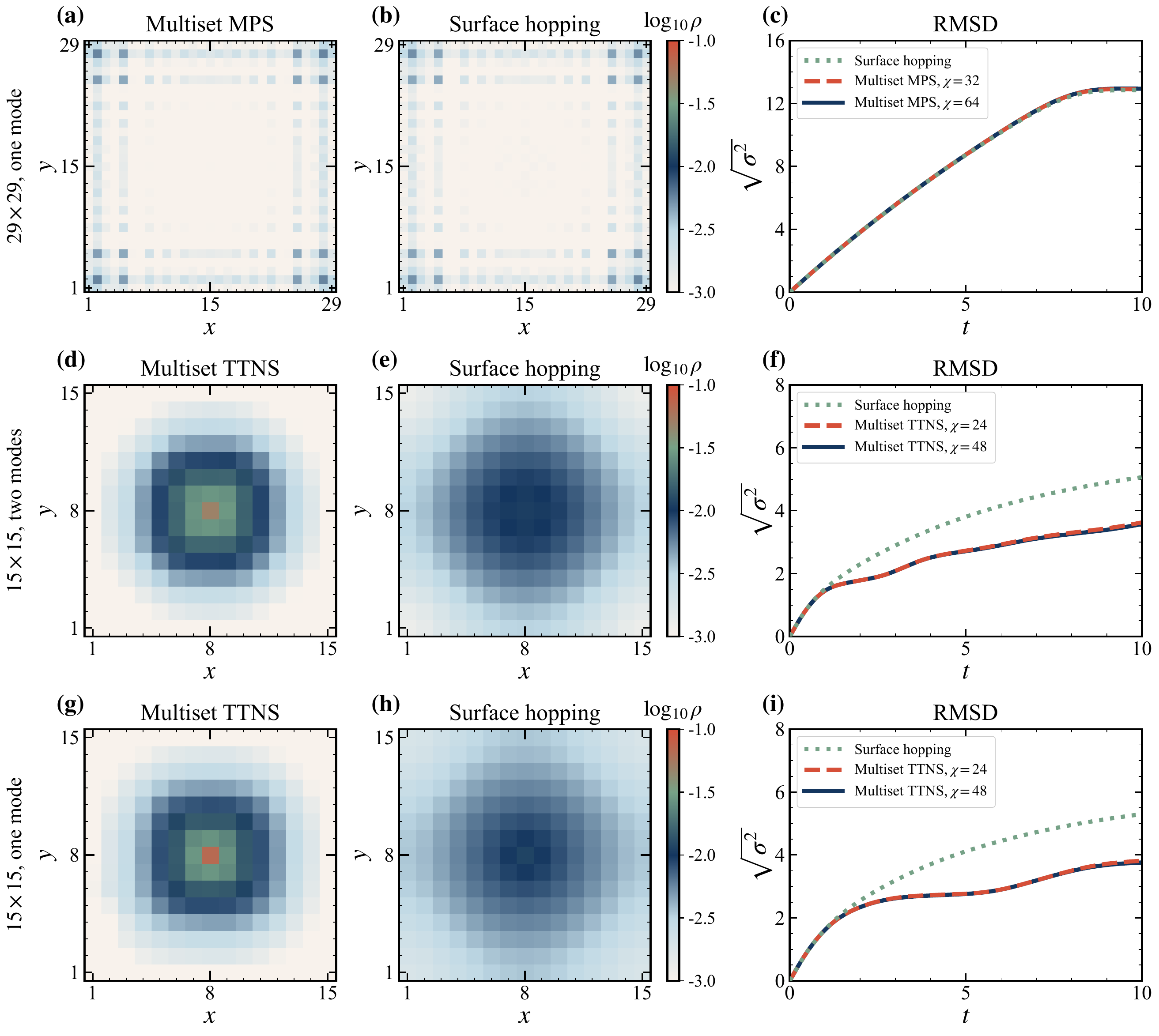}
    \caption{Exciton spreading after a central Franck--Condon excitation on open square Holstein lattices. Rows show (a--c) the weak-coupling \(29\times29\) one-mode lattice, (d--f) the strong-coupling \(15\times15\) two-mode lattice, and (g--i) the strong-coupling \(15\times15\) one-mode lattice. The first two columns show the site-resolved excitation populations obtained from multiset TDVP and surface hopping, respectively, at \(t=10\) a.u., using a logarithmic color scale. The third column shows the corresponding root mean square displacement \(\sqrt{\sigma^2}\).}
    \label{fig:holstein-large-scale}
\end{figure}

Mixed quantum–classical methods offer an efficient approach to nonadiabatic dynamics, but the accuracy of these methods in large-scale electron--vibrational systems remains insufficiently benchmarked against fully quantum dynamics. Previous studies using Simulation Package for non-Adiabatic Dynamics in Extended systems (SPADE) have investigated charge transport in two-dimensional Holstein models.\cite{Dong2025SPADE} The large-scale simulations enabled by our multi-GPU multiset framework now allow a direct comparison between surface hopping and fully quantum electron--vibrational dynamics. To assess the accuracy of surface hopping across different electron–vibrational coupling regimes, we examine the spatial spreading of the excitation at zero temperature, where nuclear fluctuations persist in the absence of thermal vibrational excitation. Our surface hopping results are averaged over \(1024\) trajectories for the weak electron-vibration coupling lattice and \(10^6\) trajectories for each strong coupling lattice, using the same model parameters and initial electronic localization as the corresponding multiset calculations. The same nearest-neighbor excitonic coupling \(J=1\) is used in all cases. Figure~\ref{fig:holstein-large-scale} compares the site populations at \(t=10\) and the root mean square displacement measured from the initial site,
\begin{equation}
    \sqrt{\sigma^2}(t)=\sqrt{
      \frac{\sum_i \rho_i(t)|\mathbf r_i-\mathbf r_0|^2}
           {\sum_i \rho_i(t)}
    }.
    \label{eq:holstein-rmsd}
\end{equation}

We first consider a \(29\times29\) lattice with one local vibrational mode coupled to each electronic site, retaining eight Fock states for each vibrational modes. With \(\omega_0=1\) and \(g=0.5\), the local reorganization energy is \(\lambda=0.25\). Following a central Franck--Condon excitation, the site populations preserve the rotational \(C_4\) symmetry of the Hamiltonian and initial state, providing a consistency check for the time evolution. In this weak-coupling regime, a pronounced ballistic component survives the local electron--vibrational interaction, as indicated by the approximately linear increase in RMSD before the wavefront reaches the lattice edges. The two multiset MPS, with \(\chi=32\) and \(64\), closely overlap over the full interval. The surface hopping calculation also reproduces the overall population pattern and the RMSD evolution, including the reduced growth as the exciton reaches the open boundaries. The corresponding population deviation in Eq.~\ref{eq:cumulative-population-error} is at the order of \(10^{-5}\) for surface hopping and \(10^{-6}\) between the two multiset MPS. These results demonstrate quantitative agreement between surface hopping and fully quantum electron--vibrational dynamics in the weak-coupling regime, extending beyond the overall excitation spreading to the spatially resolved site populations.

Despite the excellent agreement observed in the weak-coupling regime, the accuracy of surface hopping under strong electron--vibrational coupling remains less well established. To further examine surface hopping, we consider two \(15\times15\) Holstein lattices with the same nearest-neighbor electronic coupling but different vibrational configurations, as shown in Figure~\ref{fig:holstein-large-scale}(d--i). The first model contains two vibrational modes per site, with \((\omega_1,\omega_2)=(0.5,2)\) and \((g_1,g_2)=(\sqrt{8/3},\sqrt{4/3})\), retaining 32 Fock states for each of the 450 modes. The second model contains one mode per site, with \(\omega_0=1\) and \(g=2\), retaining 64 Fock states for each of the 225 modes. Both models have the same local reorganization energy \(\lambda=\sum_k\omega_k g_k^2=4\) and total Huang--Rhys factor \(S=\sum_k g_k^2=4\). By redistributing the electron–vibrational coupling among modes of different frequencies while keeping \(\lambda\) and \(S\) fixed, these two models allow us to examine how the vibrational frequency influences exciton dynamics.

Although the two strong-coupling models have identical Huang--Rhys factors and reorganization energies, their different vibrational frequency distributions give rise to distinct nuclear response timescales and electron--vibrational correlations. Previous studies of polaron transport have also demonstrated the importance of vibrational timescales beyond the reorganization energy alone.\cite{Bhattacharyya2024Mori} In our numerical simulations, the two-mode model exhibits slightly more restricted excitation spreading than the single-mode model, with RMSDs of \(3.6\) and \(3.8\) lattice spacings at \(t=10\)~a.u., respectively. In contrast to the weak-coupling regime, surface hopping fails to reproduce the excitation spreading in two strong-coupling models. Although surface hopping incorporates advanced algorithms for state tracking and decoherence corrections in extended systems, these improvements do not capture the suppression of excitation spreading induced by strong electron--vibrational coupling, as demonstrated by our fully quantum dynamics simulations.

We recognize that, despite the quantitative discrepancies observed in the strong-coupling regime, surface hopping remains a valuable approach for large-scale nonadiabatic dynamics. The independent propagation of mixed quantum--classical trajectories enables efficient parallel sampling on multiple device. For the \(29\times29\) Holstein lattice, surface hopping propagated \(1024\) trajectories to \(t=10\)~a.u. in \(2\) h of wall time, using up to 128 CPU cores on AMD EPYC 7763 nodes. By comparison, our multiset MPS calculation with \(\chi=64\) required \(301\) h on four NVIDIA A100-SXM4 GPUs (80 GB each) and two AMD EPYC 9654 processors, using one MPI rank per GPU and four CPU threads per rank. Notably, the recorded GPU memory is \(11.4\) GB per device, indicating the our calculation has not yet reached the GPU memory limitation.

Taken together, these comparisons demonstrate the value of large-scale multiset TDVP as a quantum benchmark for mixed quantum--classical dynamics. Surface hopping closely reproduces the spatially resolved excitation dynamics in the weak-coupling lattice, whereas substantial discrepancies in excitation spreading emerge in both strong-coupling models. These results highlight the importance of fully quantum electron--vibrational dynamics for assessing the accuracy and applicability of approximate transport methods in extended systems.

\subsection{Exciton dissociation on P3HT:PCBM Heterojunction}
\label{sec:p3ht-pcbm}

\begin{figure}[!htbp]
    \centering
    \includegraphics[width=\textwidth]{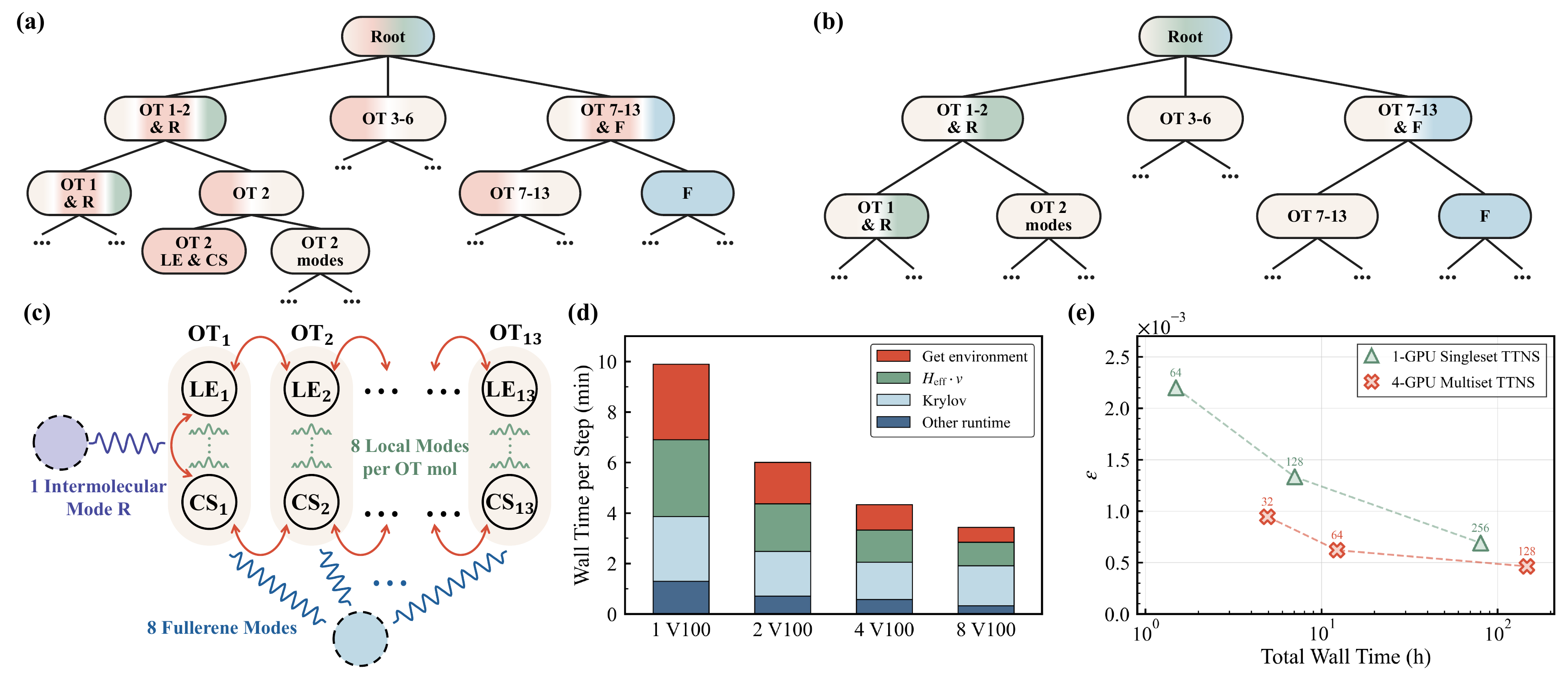}
    \caption{Multiset TTNS simulations of exciton dissociation in the P3HT:PCBM heterojunction. (a) Singleset Tree topology and (b) corresponding vibrational TTNS topology for the multiset ans{\"a}tze. (c) Schematic of the P3HT:PCBM heterojunction model. (d) Computational time analysis of multi-GPU multiset TTNS at \(\chi=64\) using one to eight NVIDIA V100 GPUs. (e) Mean cumulative population error versus total wall time for 200 fs time evolution, comparing single-GPU singleset TTNS at \(\chi=64,128,256\) (triangles) and multi-GPU multiset TTNS at \(\chi=32,64,128\) (crosses).}
    \label{fig:p3ht-pcbm}
\end{figure}

We next examine the advantages of multi-GPU multiset TDVP using a chemically motivated model of exciton dissociation at poly(3-hexylthiophene):[6,6]–phenyl–\(\mathrm{C}_{61}\)–butyric acid methyl ester (P3HT:PCBM) heterojunctions.\cite{Brey2024TransientLocalization,Dorfner2024Comparison,Lindoy2025pyTTN,Li2026MpsTtns} As illustrated in Figure~\ref{fig:p3ht-pcbm}(c), the model comprises 13 oligothiophene (OT) donor units coupled to a coarse-grained fullerene acceptor, with 26 electronic states consisting of 13 local exciton states \(|\mathrm{LE}_n\rangle\) and 13 charge-separated states \(|\mathrm{CS}_n\rangle\). Nearest-neighbor electronic couplings connect states within the LE and CS manifolds, whereas \(|\mathrm{LE_1}\rangle\) and \(|\mathrm{CS_1}\rangle\) are coupled at the interface. The 113 vibrational modes comprise eight local modes per OT unit, eight fullerene modes, and one intermolecular mode \(R\). The local OT modes modulate the corresponding LE and CS energies, the fullerene modes couple to all CS states, and the \(R\) mode modulates both the \(\mathrm{CS_1}\) energy and the \(\mathrm{LE_1}\)–\(\mathrm{CS_1}\) coupling. Using the Hamiltonian and parameters of our previous work,\cite{Li2026MpsTtns} we simulate the dynamics up to 200 fs from a Franck–Condon initial state \(|\mathrm{LE}_1\rangle\). The combination of complex electron–vibrational couplings and established high-accuracy reference data makes this system a demanding benchmark for assessing both the accuracy and computational efficiency of multiset TDVP.

We adopt the TreeX topology of Li et al.,\cite{Li2026MpsTtns} which achieved the best accuracy--cost balance among the four singleset structures tested for P3HT:PCBM Heterojunction. As shown in Figure~\ref{fig:p3ht-pcbm}(a), its three main branches group OT 1--2 with mode \(R\), OT 3--6, and OT 7--13 with the fullerene modes \(F\), while each OT subtree combines a local LE/CS electronic node and its vibrational modes. Specially, node colors distinguish the physical degrees of freedom in each subtree, with pink representing electronic states, beige OT vibrations, green \(R\) modes, and blue \(F\) modes. In the multiset ans{\"a}tze, the electronic nodes are removed from the tree topology, yielding \(N_{\mathrm e}\) independent TTNSs, each with the structure shown in Figure~\ref{fig:p3ht-pcbm}(b).

As a preliminary demonstration, we benchmark multiset TTNS at a bond dimension of \(\chi=64\) (Figure~\ref{fig:p3ht-pcbm}(d)). As the GPU count increases from one to eight, the step time decreases from 9.9 to 3.4 min, with substantial acceleration in effective Hamiltonian contractions and Krylov solver. However, the increased Krylov solver time from six to eight GPUs suggests that inter-GPU communication and synchronization overheads increasingly offset the benefits of parallelization. Notably, constructing the full electronic reduced density matrix (RDM) in the multiset ansatz requires \(N_{\mathrm e}^{2}\) pairwise overlaps between TTNSs. We therefore exclude this additional evaluation from the total wall time. For comparison, RDM evaluation accounts for only 1.5\% of the total wall time in the \(\chi=256\) singleset TTNS calculation reported by previous works.

Beyond the multi-GPU scaling analysis, we further evaluate the accuracy–cost trade-off, as shown in Figure~\ref{fig:p3ht-pcbm}(e). We adopt the cumulative population error defined in Eq.~\ref{eq:cumulative-population-error}, using the numerically exact results of Li et al.\cite{Li2026MpsTtns} as the reference. At comparable errors near \(\epsilon=7\times10^{-4}\), singleset TTNS with \(\chi=256\) requires 80 h, whereas multiset TTNS with \(\chi=64\) completes the calculation in 12.3 h, achieving a speedup of 6.5 times. The multiset TTNS calculations in this comparison were performed on four paralleled NVIDIA V100 GPUs, with more details are provided in the Supporting Information. In a nutshell, these results demonstrate the improved accuracy--cost balance of multi-GPU multiset TTNS in a complex molecular systems.

\subsection{Finite Temperature Spectra of PBI Aggregates}
\label{sec:pbi-spectra}

The vibronic spectra of perylene bisimide (PBI) aggregates reflect the interplay between excitonic coupling and intramolecular vibrations, making these systems useful benchmarks for finite-temperature quantum dynamics.\cite{Ren2018Spectra,Jiang2020FrequencySpectra,Xu2022AdaptTDVP} Here, we adopt the PBI dimer model of Ren et al.\cite{Ren2018Spectra} to calculate absorption and fluorescence spectra at 298 K. The model consists of two local electronic excited states coupled by an intermolecular excitonic interaction of \(-500\,\mathrm{cm}^{-1}\). Each molecule is coupled to ten harmonic vibrational modes with frequencies ranging from 206 to \(1628\,\mathrm{cm}^{-1}\), with five vibrational basis states retained per mode.

\begin{figure}[!t]
    \centering
    \includegraphics[width=\textwidth]{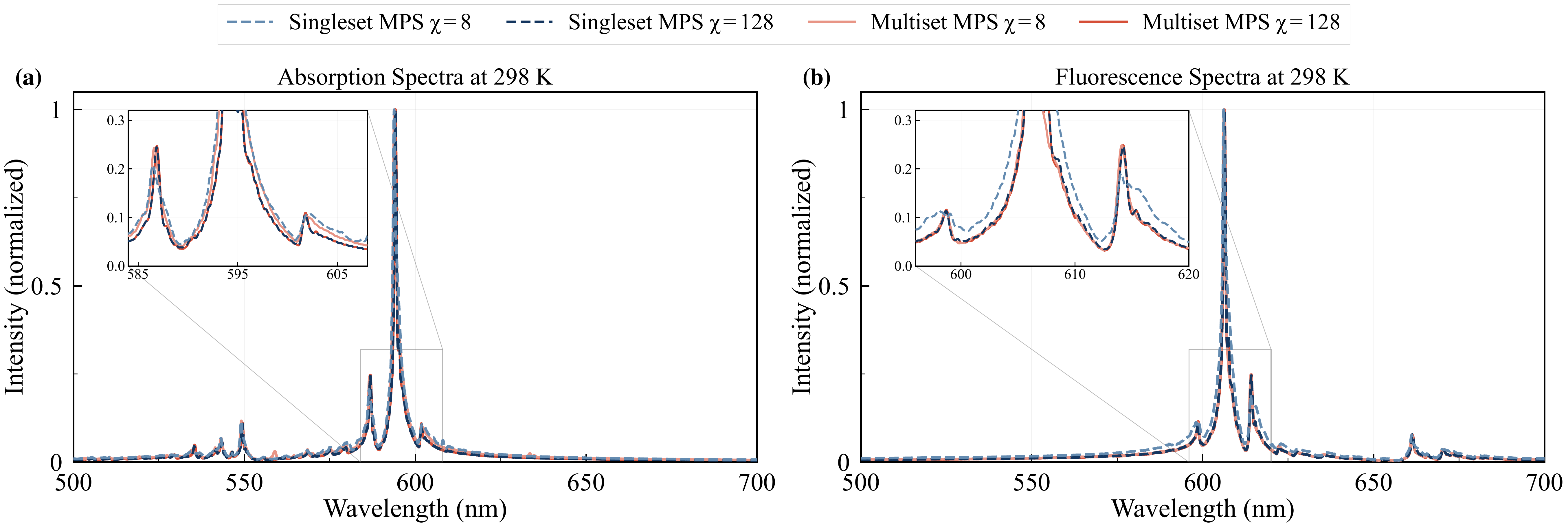}
    \caption{(a) Absorption and (b) fluorescence spectra of the PBI dimer at 298 K calculated using singleset and multiset MPS with \(\chi=8\) and 128. Insets highlight the fine structure near the main spectral peaks.}
    \label{fig:pbi-spectra}
\end{figure}

Thermal reference states are prepared by purification in the zero-exciton and one-exciton manifolds for absorption and fluorescence, respectively, following Section~\ref{sec:tfd}. Subsequently, the dipole correlation functions are then propagated to \(1\times10^5\) a.u.  with a time step of \(20\) a.u. using singleset and multiset TDVP. The spectral profiles are constructed from the real parts of the Fourier-transformed correlation functions, with frequency factors of \(\omega\) for absorption and \(\omega^3\) for fluorescence. Consequently, a Lorentzian broadening with a half width at half maximum of \(6\,\mathrm{cm}^{-1}\) is applied, and each spectrum is normalized to unit peak intensity. Figure~\ref{fig:pbi-spectra} compares the absorption and fluorescence spectra obtained with singleset and multiset MPS. Both spectra exhibit a dominant band accompanied by weaker vibronic features, as highlighted in the insets. At \(\chi=128\), the two representations agree closely in relative intensities and vibronic fine structure. However, their differences become more pronounced at \(\chi=8\). While singleset MPS captures the main spectral features, noticeable deviations emerge, particularly in the fluorescence lineshape. In contrast, multiset MPS closely reproduces the \(\chi=128\) spectra even at \(\chi=8\), highlighting its ability to achieve higher accuracy at a smaller bond dimension. Taken together, these results extend the application of multiset TDVP from population dynamics to thermal optical response and demonstrate its practical applicability to optical spectroscopy of molecular aggregates.

\section{Conclusion and Outlook}
\label{sec:conclusion}

We have developed a scalable multi-GPU framework for multiset MPS and TTNS quantum dynamics within the Renormalizer package. Compared with previous GPU-accelerated TDVP implementations, we distribute tensor contractions and Krylov vectors across multiple GPUs, reducing the memory demand and computational workload on each GPU. Our large-scale fully quantum simulations provide a stringent test of mixed quantum-classical methods, whose applicability and computational efficiency are receiving increasing attention nowadays. Furthermore, the application of the multiset ans{\"a}tze to exciton dissociation in a P3HT:PCBM heterojunction and its extension to finite-temperature spectroscopy of aggregate demonstrate the versatility of our works. While the overhead of inter-GPU communication is not immediate eliminated, we expect that the improvement will become significant for moderate and large-scale quantum systems.

More broadly, the parallelization strategies introduced here are not restricted to the multiset ans{\"a}tze and can be adapted to other tensor-network formulations of quantum dynamics. A promising extension of the present multi-GPU framework is its application to tensor network formulations of hierarchical equations of motion (HEOM). Efficiently handling couplings among auxiliary density operators and ensuring stable non-Hermitian time propagation present nontrivial challenges for extending the current multi-GPU parallelization strategies. Building on our multi-GPU parallelization strategies and recent methodological advances in HEOM,\cite{Chen2025TTNHEOM,RodriguezBetancourt2026TENSO,Javanmard2026BiorthogonalTDVP,Li2026GaugeHEOM} our primary plan is to generalize the multi-GPU framework and explore the accessible system sizes for non-Markovian open quantum dynamics in future work.

\section*{Acknowledgement}

We thank Zhigang Shuai, Linjun Wang, and Yaxin Wang for helpful discussions. This work was supported by the National Natural Science Foundation of China (grant nos. 22595405 and  22573088), the Guangdong Basic and Applied Basic Research Foundation (grant nos. 2026A1515010545), the Guangdong Provincical Quantum Science Strategic Initiative (grant nos. GDZX2503001), the Shenzhen Science and Technology Program (grant nos. KQTD20240729102028011 and JCYJ20250604140259001), and the Guangdong Basic Research Center of Excellence for Aggregate Science.

\section*{Data and Code Availability}

The source code for multi-GPU multiset tensor network is publicly available on GitHub 
\href{https://github.com/Relinquish0/Renormalizer/tree/feature/R178-multiset/code_multiGPUms}{https://github.com/Relinquish0/Renormalizer}.

\section*{Supporting Information}

The Supporting Information, appended after the references, contains Hamiltonian-guided initial bond expansion, distributed Lanczos--Krylov propagation, the symmetric multiset projector-splitting sweep, and state-resolved population dynamics of the P3HT:PCBM heterojunction.

\bibliography{refs}

\end{document}


\section{Hamiltonian-Guided Initial Bond Expansion}
\label{sec:si-hamiltonian-guided-expansion}

Because one-site TDVP confines the trajectory to the tangent space of a fixed-rank MPS manifold, its bond dimensions do not increase during propagation. A product initial state consequently lacks the variational directions needed to represent correlations generated immediately by the Hamiltonian. We instead initialize the enlarged manifold with directions generated by the block Hamiltonian.

For the conditional MPS associated with electronic state \(\alpha\), we first construct the expansion vector \(\ket{v_0^\alpha}=\ket{\Phi^\alpha}+\sum_{\beta\neq\alpha}\hat H^{\alpha\beta}\ket{\Phi^\beta}\). Each off-diagonal block \(\hat H^{\alpha\beta}\) maps the vibrational wavepacket conditioned on state \(\beta\) into the vibrational space associated with state \(\alpha\). The vector \(\ket{v_0^\alpha}\) therefore contains the directions through which amplitude can initially enter the \(\alpha\)-conditioned MPS.

Using the MPS addition rule, the initial conditional MPS is expanded as
\begin{equation}
    \ket{\Phi_{\mathrm{exp}}^\alpha}
    =
    \ket{\Phi^\alpha}
    +
    \epsilon_{\mathrm{exp}}
    \ket{\Xi^\alpha},
    \label{eq:si-expanded-initial-component}
\end{equation}
where \(\ket{\Xi^\alpha}\) is constructed from the Hamiltonian Krylov space
\begin{equation}
    \mathcal K_m^\alpha
    =
    \operatorname{span}
    \left\{
        \ket{v_0^\alpha},
        \hat H^{\alpha\alpha}\ket{v_0^\alpha},
        \ldots,
        \left(\hat H^{\alpha\alpha}\right)^{m-1}
        \ket{v_0^\alpha}
    \right\}.
    \label{eq:si-multiset-expansion-krylov-space}
\end{equation}
The Krylov vectors are orthogonalized against the original MPS before being assembled into \(\ket{\Xi^\alpha}\). The scalar \(\epsilon_{\mathrm{exp}}\) is chosen sufficiently small that the physical initial state is unchanged to leading order. After the MPS addition, the expanded state is rescaled to retain the norm of the original \(\alpha\)-conditioned MPS. The expansion is applied only at initialization; subsequent one-site TDVP propagation uses the resulting fixed bond dimensions.

\section{Distributed Krylov Propagation}
\label{sec:si-distributed-krylov}

Algorithm~\ref{alg:si-distributed-krylov} summarizes the distributed solver used for each one-site and zero-site TDVP propagation. Rank \(p\) stores only the rows of the canonical-center vector and Krylov basis associated with its assigned multiset components. The effective-Hamiltonian action returns the same local row range. For a Hermitian effective Hamiltonian \(H\), the algorithm evaluates the local blocks of \(y=\exp(zH)x\), where the complex coefficient \(z\) includes the sign and duration of the local propagation.

The superscript \((p)\) denotes a local vector block. The global inner product \(\langle u,v\rangle_{\mathrm g}\) is the sum of the local inner products over all ranks, evaluated by \(\AllreduceSum\). The global norm \(\|u\|_{\mathrm g}\) is the square root of the real part of the global self-inner product. Thus, every rank obtains the same Lanczos coefficients and constructs the same tridiagonal projected matrix \(T_m\), with diagonal entries \(a_1,\ldots,a_m\) and off-diagonal entries \(b_1,\ldots,b_{m-1}\). The matrix \(V_m^{(p)}\) contains the first \(m\) local basis vectors as columns, \(e_1\) is the first unit vector, and \(\eta\) is the global input norm.

The Krylov space is extended by \(k\) basis vectors between convergence checks. Successive checked approximations are compared using \texttt{allclose}, with absolute and relative tolerances \texttt{atol} and \texttt{rtol}. The local Boolean results are combined by a global logical AND reduction. The first checked approximation initializes \(y_{\mathrm{old}}\); it cannot establish convergence by comparison. If the residual norm falls below the Lanczos breakdown tolerance \(\delta_{\mathrm{bd}}\), the current approximation is returned immediately.

The full Krylov basis is never assembled on a single GPU. The replicated quantities consist only of the initial norm, the Lanczos coefficients, the convergence flag, and the small projected matrix. The local basis blocks together represent one globally orthonormal Krylov basis. In the implementation, basis storage is extended in allocation blocks as needed; this memory-management detail does not alter the recurrence.

\clearpage
\begin{algorithm}[!htbp]
\caption{Distributed Lanczos--Krylov propagation. Each rank executes the same steps on its local vector blocks. Inner products and norms are evaluated globally.}
\label{alg:si-distributed-krylov}
\small
\setstretch{1.1}
\begin{algorithmic}[1]
\Require Distributed vector \(x\), effective-Hamiltonian action \(v\mapsto Hv\), coefficient \(z\), basis-extension block size \(k\), \texttt{allclose} tolerances \texttt{atol}, \texttt{rtol}, and breakdown tolerance \(\delta_{\mathrm{bd}}\).
\Ensure Local block \(y^{(p)}\) and Krylov dimension \(m\).
\State \(\eta\gets\|x\|_{\mathrm g}\)
\If{\(\eta=0\)}
    \State \Return \((0,0)\)
\EndIf
\State \(v_1^{(p)}\gets x^{(p)}/\eta\), \(v_0^{(p)}\gets0\), \(b_0\gets0\), \(m\gets0\)
\State Mark \(y_{\mathrm{old}}^{(p)}\) as unavailable.
\While{true}
    \State \(m\gets m+1\)
    \State Exchange the vector blocks required for \((Hv_m)^{(p)}\).
    \State \(w^{(p)}\gets(Hv_m)^{(p)}-b_{m-1}v_{m-1}^{(p)}\)
    \State \(a_m\gets\operatorname{Re}\langle v_m,w\rangle_{\mathrm g}\)
    \State \(w^{(p)}\gets w^{(p)}-a_mv_m^{(p)}\), \quad \(b_m\gets\|w\|_{\mathrm g}\)
    \If{\(b_m\leq\delta_{\mathrm{bd}}\) or \(m\bmod k=0\)}
        \State Construct \(T_m\) from \(a_1,\ldots,a_m\) and \(b_1,\ldots,b_{m-1}\).
        \State \(y_m^{(p)}\gets\eta V_m^{(p)}\exp(zT_m)e_1\)
        \If{\(b_m\leq\delta_{\mathrm{bd}}\)}
            \State \Return \((y_m^{(p)},m)\)
        \EndIf
        \If{a previous approximation is available}
            \State \(c_p\gets\operatorname{allclose}(y_m^{(p)},y_{\mathrm{old}}^{(p)},\texttt{atol=atol},\texttt{rtol=rtol})\)
            \State \(c\gets\operatorname{Allreduce}_{\mathrm{AND}}(c_p)\)
            \If{\(c\)}
                \State \Return \((y_m^{(p)},m)\)
            \EndIf
        \EndIf
        \State \(y_{\mathrm{old}}^{(p)}\gets y_m^{(p)}\)
    \EndIf
    \State \(v_{m+1}^{(p)}\gets w^{(p)}/b_m\); retain \(v_{m+1}^{(p)}\) in the local basis.
\EndWhile
\end{algorithmic}
\end{algorithm}

\clearpage
\section{Symmetric Multiset Projector-Splitting Sweep}
\label{sec:si-multiset-sweep}

Algorithm~\ref{alg:si-multiset-sweep} gives one real-time step of the one-site multiset MPS integrator. It follows Algorithm~1 in the Supporting Material of Kloss et al.,\cite{Kloss2019Multiset} using the notation of Sec.~2.2 of the main text. The multi-GPU implementation retains this symmetric projector-splitting sweep and distributes its tensor contractions and matrix-exponential actions across GPUs.

The vectors \(C_j\) and \(D_j\) collect the site-center coefficients and bond matrices for all electronic states. The corresponding Hamiltonian actions are
\begin{align*}
    (H_{\mathrm{eff},j}C_j)^\alpha&=\sum_\beta H_{\mathrm{eff},j}^{\alpha\beta}C_j^\beta,\\
    (K_{\mathrm{eff},j}D_j)^\alpha&=\sum_\beta K_{\mathrm{eff},j}^{\alpha\beta}D_j^\beta.
\end{align*}
Every exponential in Algorithm~\ref{alg:si-multiset-sweep} therefore acts jointly on all electronic states. QR factorizations and absorption into a neighboring tensor are carried out separately for each electronic state. In a left-to-right center shift, \(C_j^\alpha\) is unfolded with \((l,\sigma_j)\) as the row index and factorized as \(A_{L,j}^\alpha D_j^\alpha\). For a right-to-left shift, a QR decomposition of the conjugate-transposed unfolding yields \(D_j^\alpha A_{R,j+1}^\alpha\). The effective Hamiltonian is evaluated in the current canonical bases and held fixed during each local exponential action. Updated environments are used after every center shift.

\clearpage
\begin{algorithm}[htbp]
\caption{Symmetric one-site multiset MPS propagation. Factorization and absorption steps apply to each electronic state \(\alpha\); exponential actions propagate the coupled set of electronic components.}
\label{alg:si-multiset-sweep}
\small
\setstretch{1.1}
\begin{algorithmic}[1]
\Require Multiset MPS with canonical center at site 1, Hamiltonian MPOs \(\{\hat H^{\alpha\beta}\}\), time step \(\tau\), and fixed bond dimensions.
\Ensure Multiset MPS at \(t+\tau\), with the canonical center at site 1.
\State Construct the right contraction environments.
\For{\(j=1,\ldots,L-1\)}
    \State \(C_j\gets\exp(-i\tau H_{\mathrm{eff},j}/2)C_j\)
    \State Obtain \(A_{L,j}^\alpha,D_j^\alpha\) from \(C_j^\alpha=A_{L,j}^\alpha D_j^\alpha\) by QR; update the left environments.
    \State \(D_j\gets\exp(+i\tau K_{\mathrm{eff},j}/2)D_j\)
    \State \(C_{j+1}^\alpha\gets D_j^\alpha A_{R,j+1}^\alpha\)
\EndFor
\State \(C_L\gets\exp(-i\tau H_{\mathrm{eff},L})C_L\)
\For{\(j=L-1,\ldots,1\)}
    \State Obtain \(D_j^\alpha,A_{R,j+1}^\alpha\) from \(C_{j+1}^\alpha=D_j^\alpha A_{R,j+1}^\alpha\) by QR; update the right environments.
    \State \(D_j\gets\exp(+i\tau K_{\mathrm{eff},j}/2)D_j\)
    \State \(C_j^\alpha\gets A_{L,j}^\alpha D_j^\alpha\)
    \State \(C_j\gets\exp(-i\tau H_{\mathrm{eff},j}/2)C_j\)
\EndFor
\end{algorithmic}
\end{algorithm}

The positive sign in the bond exponential implements backward integration of the bond equation in Sec.~2.2. The update at site \(L\) combines two consecutive half-step site propagations. The ascending sweep and its reverse form a symmetric integrator with local splitting error \(O(\tau^3)\). Each exponential is evaluated by Algorithm~\ref{alg:si-distributed-krylov}, with \(z=-i\tau/2\) for a site half step, \(z=+i\tau/2\) for a bond half step, and \(z=-i\tau\) for the last-site update. No singular-value truncation is performed during the sweep.

For multiset TTNS, the same electronic-state coupling is retained while the chain traversal is replaced by a traversal of the chosen tree and its reverse. Node-center propagation alternates with backward propagation on the traversed edge. The canonical bases and effective Hamiltonians are formed from the subtrees incident on that node or edge. Algorithm~\ref{alg:si-multiset-sweep} specifies the chain case.

\clearpage
\section{Numerical Results of P3HT:PCBM Heterojunction}

\begin{figure}[!htbp]
    \centering
    \includegraphics[width=\textwidth,height=0.80\textheight,keepaspectratio]{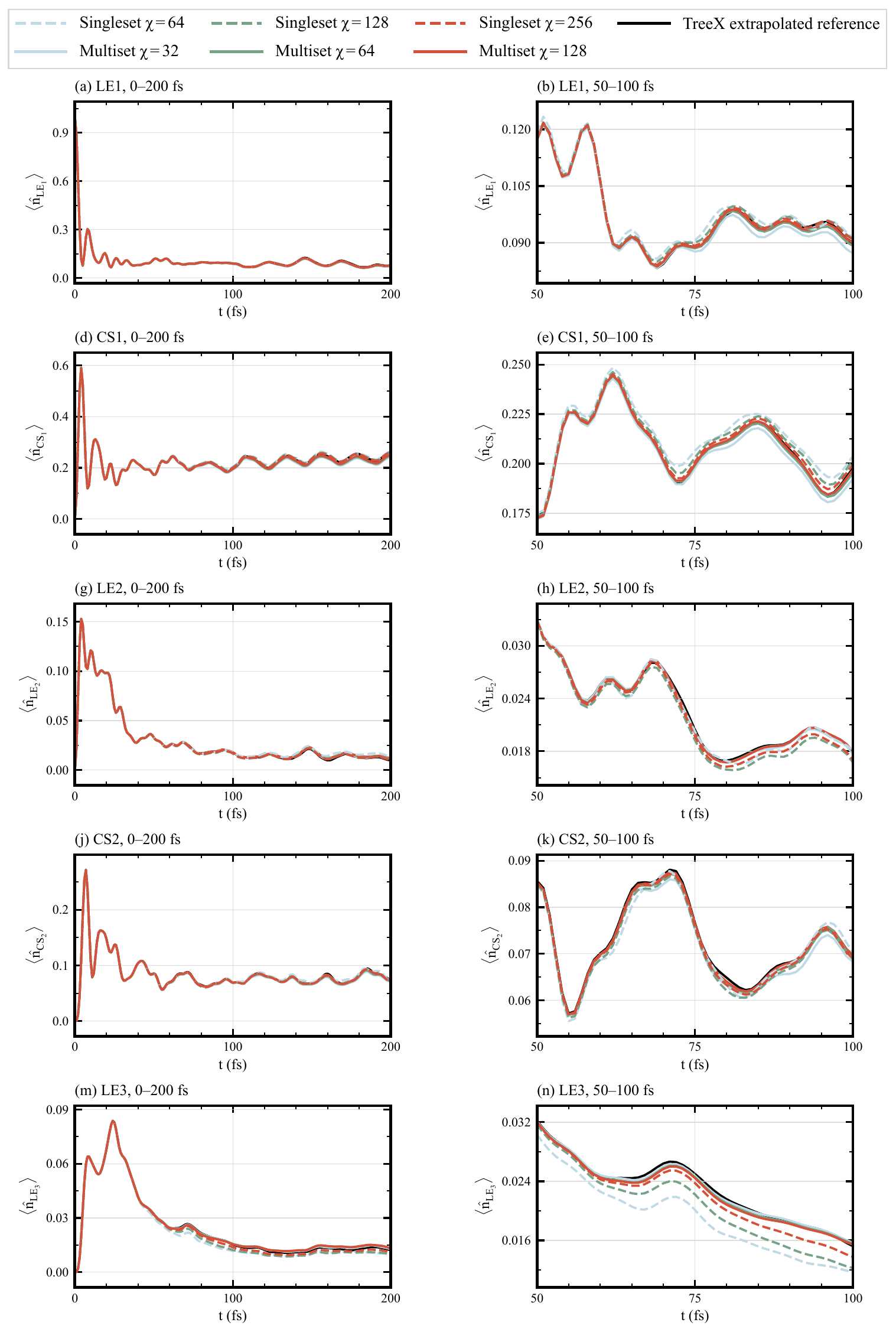}
    \caption{Electronic populations over 0--200 fs (left) and 50--100 fs (right), where singleset TTNS deviates from the reference.}
    \label{fig:si-populations-1}
\end{figure}
\clearpage

\begin{figure}[!htbp]
    \centering
    \includegraphics[width=\textwidth,height=0.89\textheight,keepaspectratio]{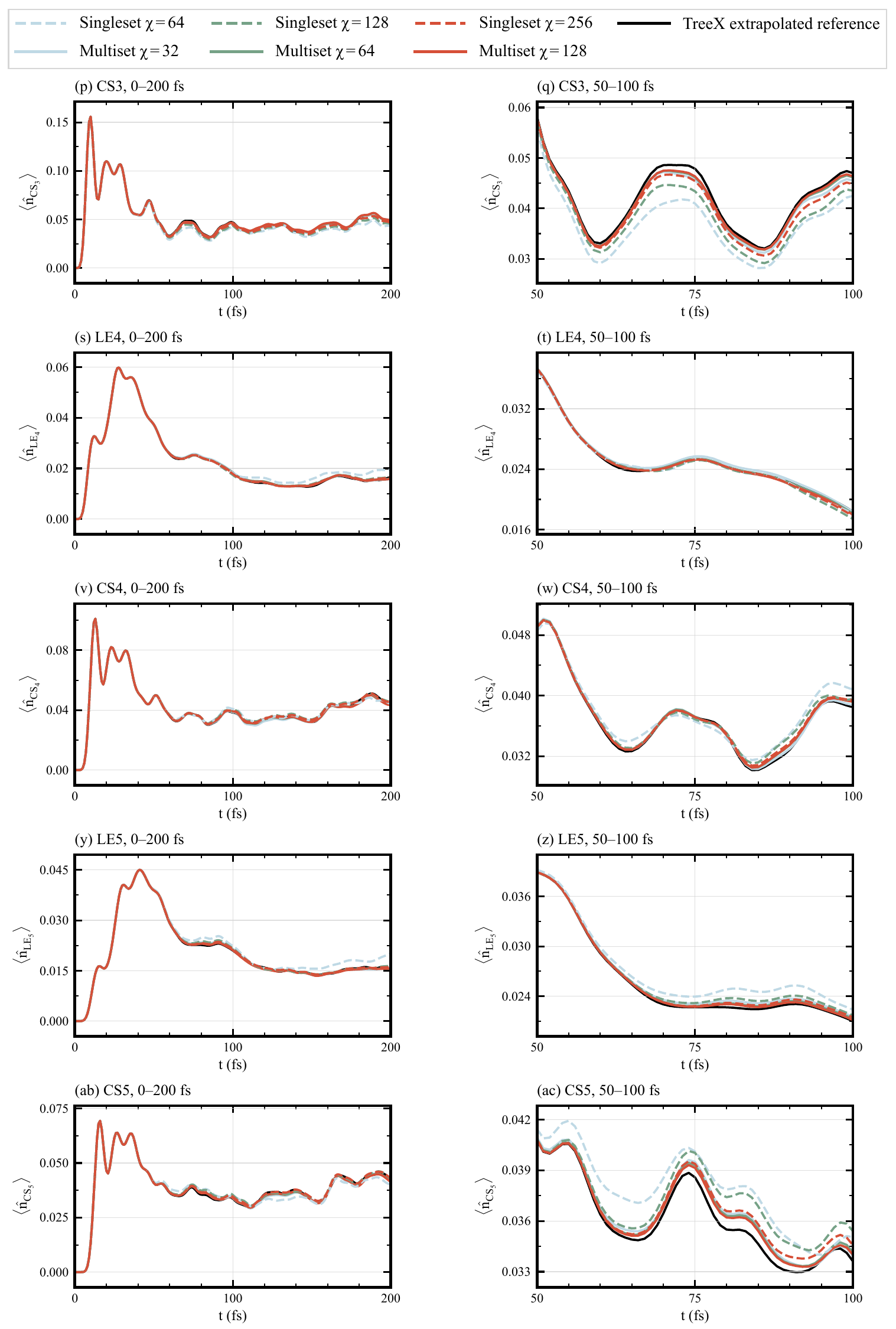}
    \caption{Electronic populations over 0--200 fs (left) and 50--100 fs (right), where singleset TTNS deviates from the reference.}
    \label{fig:si-populations-2}
\end{figure}
\clearpage

\begin{figure}[!htbp]
    \centering
    \includegraphics[width=\textwidth,height=0.89\textheight,keepaspectratio]{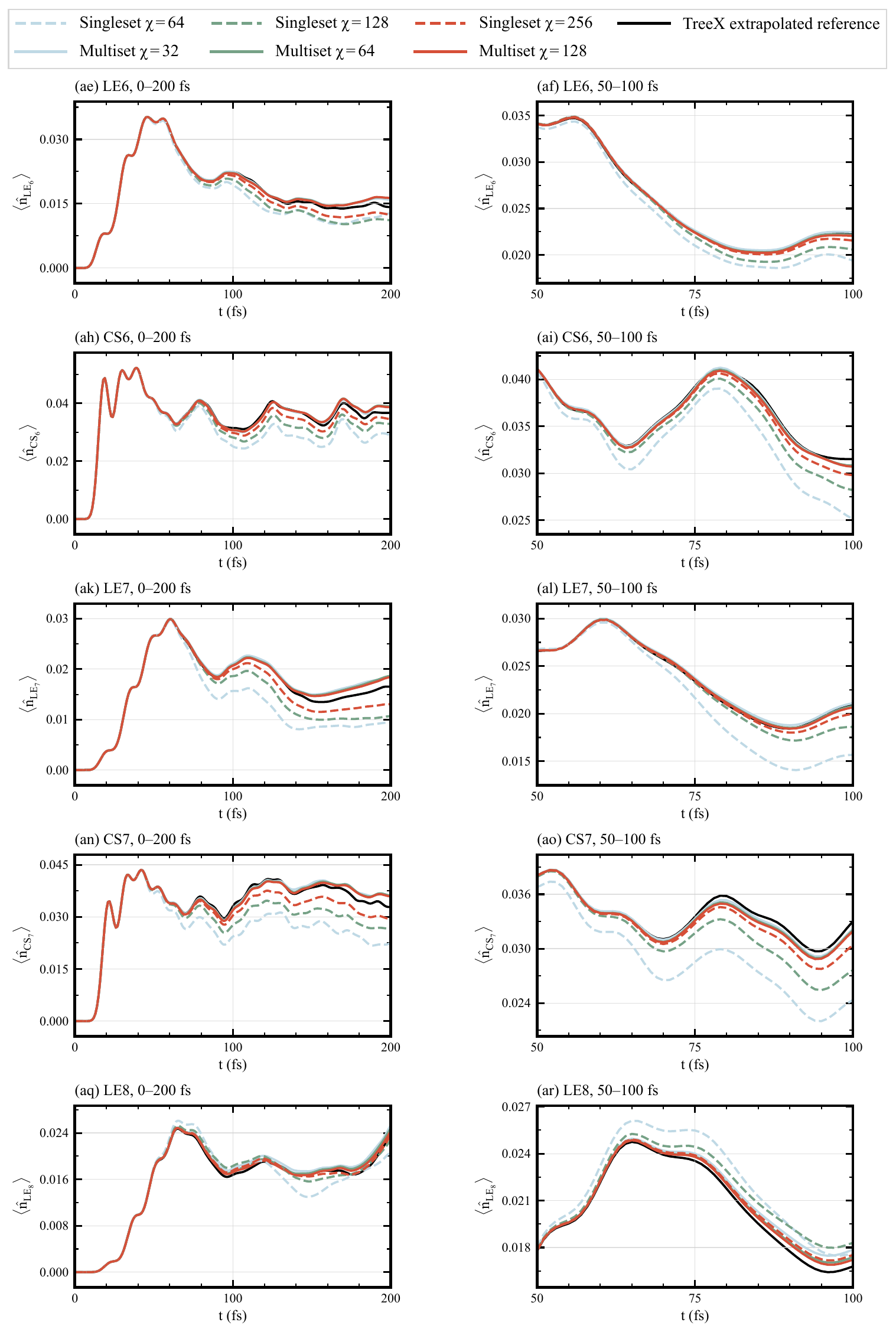}
    \caption{Electronic populations over 0--200 fs (left) and 50--100 fs (right), where singleset TTNS deviates from the reference.}
    \label{fig:si-populations-3}
\end{figure}
\clearpage

\begin{figure}[!htbp]
    \centering
    \includegraphics[width=\textwidth,height=0.89\textheight,keepaspectratio]{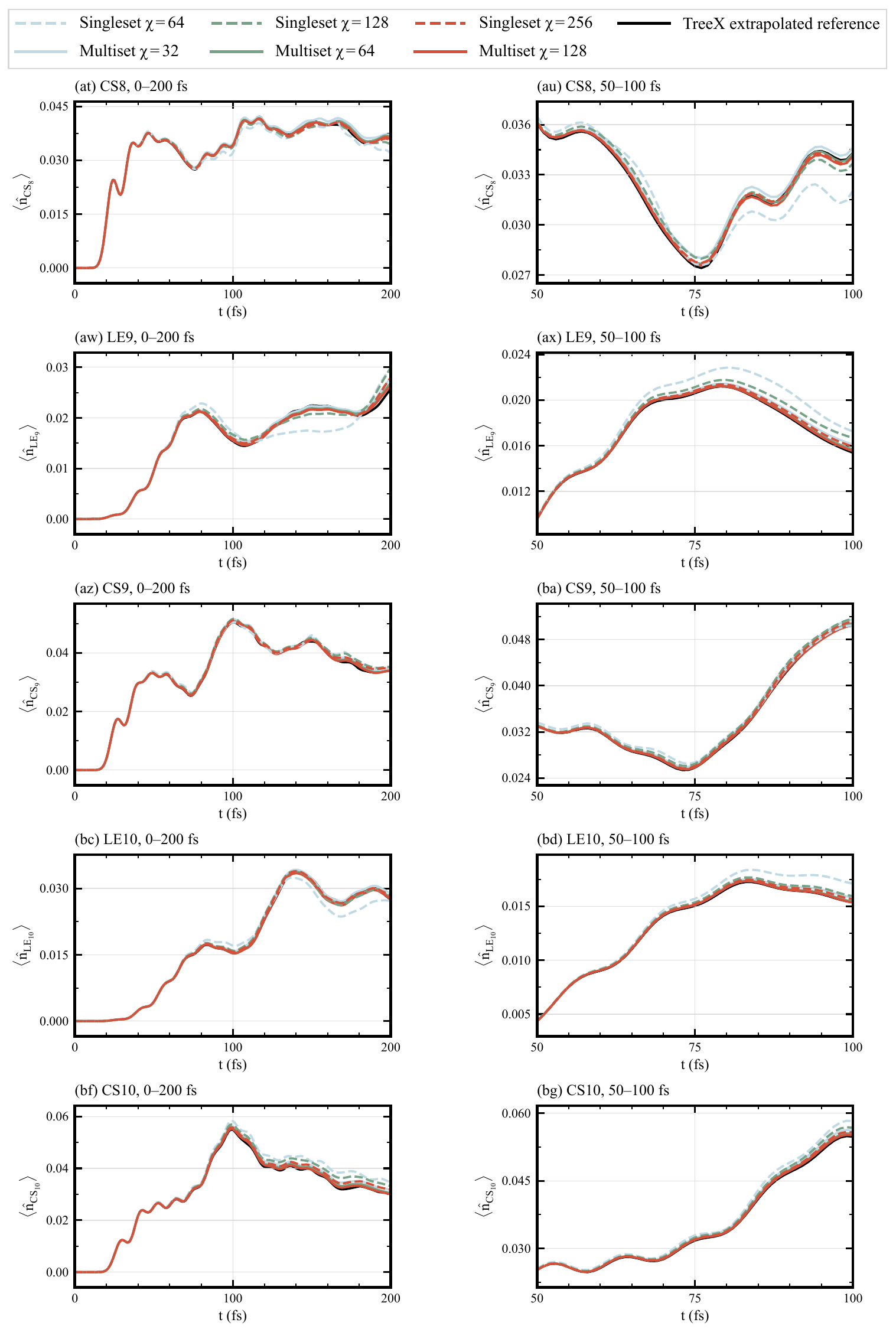}
    \caption{Electronic populations over 0--200 fs (left) and 50--100 fs (right), where singleset TTNS deviates from the reference.}
    \label{fig:si-populations-4}
\end{figure}
\clearpage

\begin{figure}[!htbp]
    \centering
    \includegraphics[width=\textwidth,height=0.89\textheight,keepaspectratio]{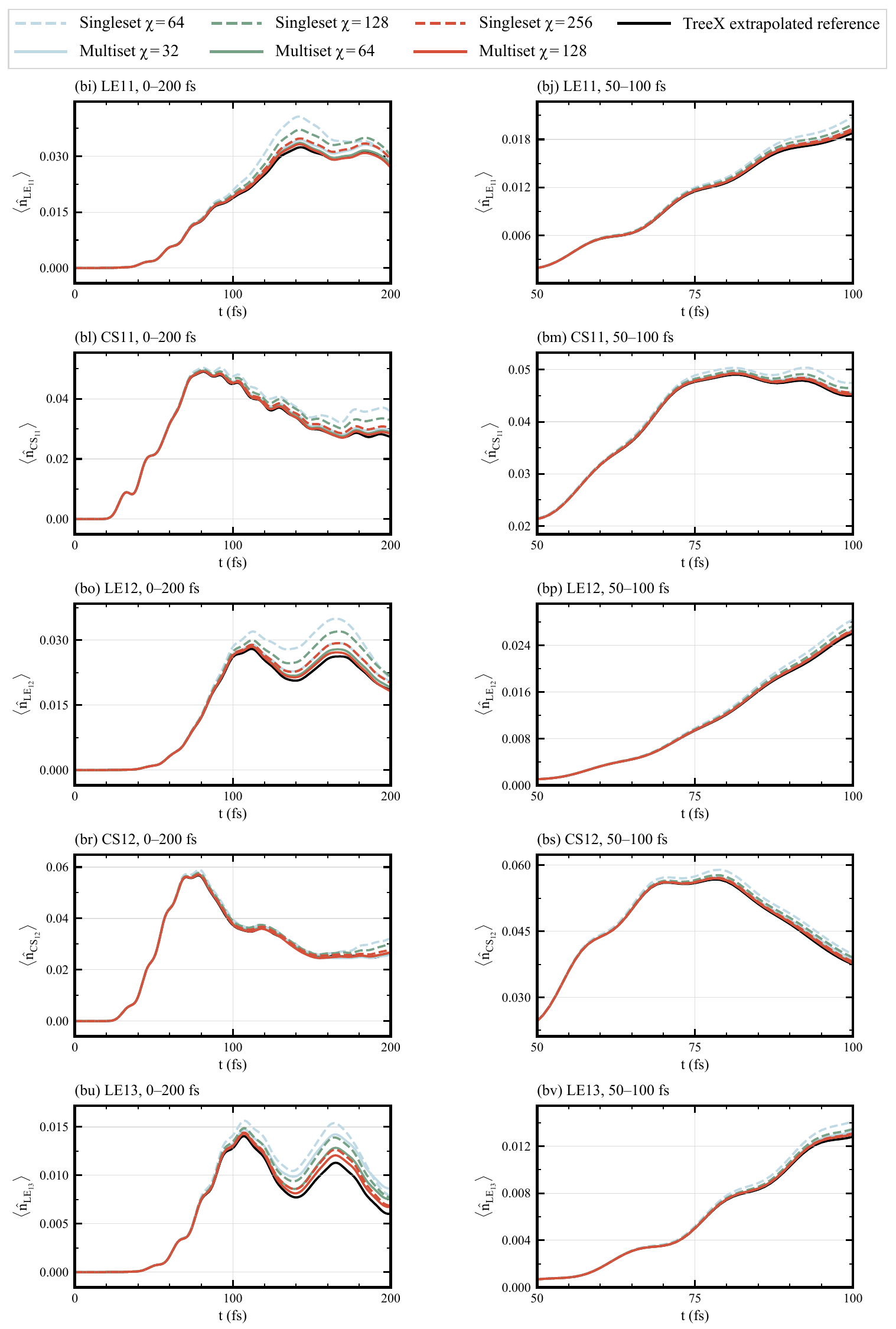}
    \caption{Electronic populations over 0--200 fs (left) and 50--100 fs (right), where singleset TTNS deviates from the reference.}
    \label{fig:si-populations-5}
\end{figure}
\clearpage

\begin{figure}[!htbp]
    \centering
    \includegraphics[width=\textwidth,height=0.89\textheight,keepaspectratio]{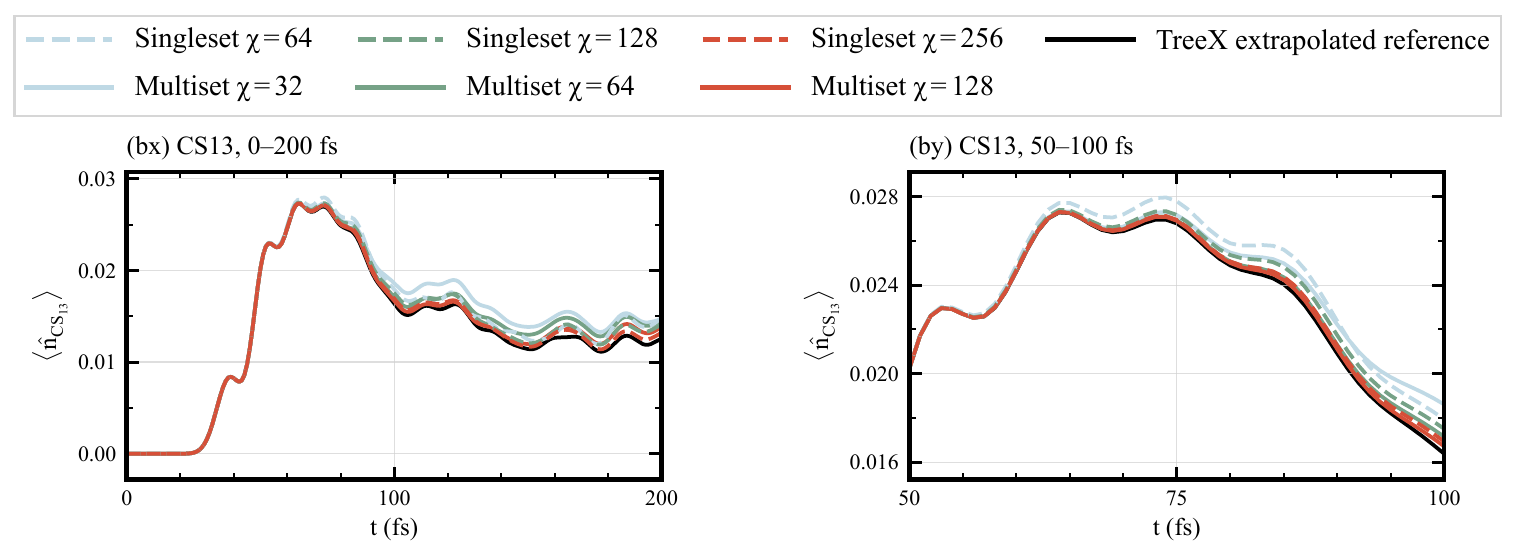}
    \caption{Electronic populations over 0--200 fs (left) and 50--100 fs (right), where singleset TTNS deviates from the reference.}
    \label{fig:si-populations-6}
\end{figure}
\clearpage

\bibliography{refs}